\documentclass[a4paper, 12pt]{article}

\pdfoutput = 1

\usepackage{styleBuding}
\usepackage{bm}
\usepackage{color}
\usepackage[usenames,dvipsnames,svgnames,table]{xcolor}
\usepackage{amsmath}
\usepackage{amssymb}
\usepackage{graphicx}
\usepackage{slashed}
\usepackage{soul}
\usepackage{multirow}
\usepackage{subfigure}
\usepackage{siunitx} 
\usepackage{url} 
\usepackage[utf8]{inputenc}

\newcommand{\rom}[1]{\uppercase\expandafter{\romannumeral #1\relax}}

\let\jnfont=\rm
\def\NPB#1,{{\jnfont Nucl.\ Phys.\ B }{\bf #1},}
\def\PLB#1,{{\jnfont Phys.\ Lett.\ B }{\bf #1},}
\def\EPJC#1,{{\jnfont Eur.\ Phys.\ Jour.\ C }{\bf #1},}
\def\PRD#1,{{\jnfont Phys.\ Rev.\ D }{\bf #1},}
\def\PRL#1,{{\jnfont Phys.\ Rev.\ Lett.\ }{\bf #1},}
\def\MPLA#1,{{\jnfont Mod.\ Phys.\ Lett.\ A }{\bf #1},}
\def\JPG#1,{{\jnfont J.\ Phys.\ G}{\bf #1},}
\def\CTP#1,{{\jnfont Commun.\ Theor.\ Phys.\ }{\bf #1},}
\def\ZPC#1,{{\jnfont Z.\ Phys.\ C }{\bf #1},}
\def\JHEP#1,{{\jnfont JHEP \ }{\bf #1},}

\title{Singlino-dominated dark matter in general NMSSM}
\author{Junjie Cao$^a$, Demin Li$^a$, Jingwei Lian$^a$, Yuanfang Yue$^a$, and Haijing Zhou$^a$}
\affiliation{ $^a$ Department of Physics, Henan Normal University, Xinxiang 453007, China}
\emailAdd{junjiec@alumni.itp.ac.cn}
\emailAdd{deminli@foxmail.com}
\emailAdd{ljwfly@hotmail.com}
\emailAdd{yuanfang405@gmail.com}
\emailAdd{zhouhaijing0622@163.com}
\abstract{The general Next-to-Minimal Supersymmetric Standard Model (NMSSM) describes the singlino-dominated dark-matter (DM) property by four independent parameters: singlet-doublet Higgs coupling coefficient $\lambda$,
Higgsino mass $\mu_{tot}$, DM mass $m_{\tilde{\chi}_1^0}$, and singlet Higgs self-coupling coefficient $\kappa$.
The first three parameters strongly influence the DM-nucleon scattering rate, while $\kappa$ usually affects the scattering only slightly.
This characteristic implies that singlet-dominated particles may form a secluded DM sector. Under such a theoretical structure, the DM achieves the correct abundance by annihilating into a pair of singlet-dominated Higgs bosons by adjusting $\kappa$'s value. Its scattering with nucleons is suppressed when $\lambda v/\mu_{tot}$ is small. This speculation is verified by sophisticated scanning of the theory's parameter space with various experiment constraints considered. In addition, the Bayesian evidence of the general NMSSM and that of $Z_3$-NMSSM is computed. It is found that, at the cost of introducing one additional parameter, the former is approximately $3.3 \times 10^3$ times the latter. This result corresponds to Jeffrey's scale of 8.05 and implies that the considered experiments strongly prefer the general NMSSM to the $Z_3$-NMSSM.

}

\begin{document}
    \maketitle
    \flushbottom
\section{Introduction}

With the smooth development of various dark-matter (DM) detection experiments in the past decades, DM research has gradually become a hot topic. Massive Weakly Interacting Particles (WIMPs) are considered the most promising DM candidates because they can naturally predict the abundance measured by the Planck experiment in~\cite{Ade:2015xua, Aghanim:2018eyx}. This phenomenon is called the WIMP Miracle in the literature~\cite{Jungman1995Supersymmetric,Bertone:2004pz}. One distinctive feature of the candidates is that the spin-independent (SI) cross-section of their scattering with nucleons is approximately $10^{-45} {\rm cm^2}$ since they are usually supposed to couple to the Standard Model (SM) particles through weak interactions~\cite{Baum:2017enm}. Nevertheless, the DM direct detection experiments give a different answer; that is, so far, the XENON-1T and PandaX-II experiments have restricted the cross-section below the order of $10^{-46} {\rm cm^2}$~\cite{Aprile:2018dbl,Wang:2020coa,Cui:2017nnn}, which implies that the interaction between DM and nucleons is at most feeble~\cite{Cao:2018rix}. This fact reveals a general conclusion that simple WIMP theories face significant experimental challenges.

In popular supersymmetric theories, the lightest neutralino, $\tilde{\chi}_1^0$, as the lightest supersymmetric particle (LSP), is a stable WIMP in theories' natural parameter space and usually acts as a DM candidate. It affects both the signal of supersymmetric particles at colliders and the evolution of the universe, so it has been the focus of DM physics over the past several decades~\cite{Jungman1995Supersymmetric}. With the increasing sensitivity of DM direct detection experiments, this candidate is becoming increasingly more difficult to be consistent with both the detection experiments and the abundance naturally~\cite{Cao:2018rix}. Taking the Minimal Supersymmetric Standard Model (MSSM) as an example, the neutralino with $m_{\tilde{\chi}_1^0} \lesssim 1~{\rm TeV}$ must be Bino-dominated in its component to produce correct relic abundance. In this case, the SI and spin-dependent (SD) cross-sections rely on Higgsino mass in different ways. If the Higgsinos are moderately light, at least one of them would be considerable in size~\cite{Cao:2019qng}. As a result, the Higgsinos must be heavier than approximately 300 GeV after considering current XENON-1T experimental constraints on the cross-sections and approximately 600 GeV supposing no DM signal detected in the future LZ experiments~\cite{Cao:2019qng}. Additionally, a global fit of the MSSM to various experimental data known in 2017 preferred the Higgsinos to be heavier than approximately $350~{\rm GeV}$ at a $95\%$ confidence level~\cite{Bagnaschi:2017tru}. This conclusion should be significantly improved since both DM detection experiments and the LHC search for supersymmetry have progressed rapidly in recent years. These facts reveal that the theory already has undergone a relatively significant fine-tuning to predict $Z$ boson mass. With further development of DM detection experiments, the tuning will become even more significant.

The dilemma of the MSSM inspired our study of the Next-to-Minimal Supersymmetric Standard Model (NMSSM)~\cite{Ellwanger:2009dp}, which extends the MSSM by one gauge-singlet superfield and is another popular realization of supersymmetry, and to consider a Singlino-dominated neutralino as a DM candidate. The theory introduces the inputs $\lambda$ and $\kappa$ to parameterize singlet-doublet Higgs interactions and singlet Higgs self-interactions. Thus, the couplings between the DM and nucleus depend on the Higgsino composition of the DM, which, for fixed DM mass, is determined by the dimensionless parameter $\lambda$ and the Higgsino mass $\mu_{tot}$ (see discussions in this paper). When $\lambda \gtrsim 0.3$ and $\mu_{tot} \lesssim 500~{\rm GeV}$, the scattering cross-sections of the DM and nucleus tend to be too large to be consistent with the XENON-1T constraints under the premise of the correct DM abundance~\cite{Cao:2016cnv}. This situation remains valid for any DM mass when the constraints from the LHC search for supersymmetry are included~\cite{Cao:2018rix,cao:2021new-work}, and it has a great impact on DM abundance~\cite{Cao:2019qng}. For example, the annihilations $\tilde{\chi}_1^0 \tilde{\chi}_1^0 \to t \bar{t}, h A_s, h_s A_s$ were believed to play crucial roles in determining the abundance~\cite{Baum:2017enm,Cao:2015loa}, where $t$, $h$, $h_s$, and $A_s$ denote the top quark, SM-like Higgs boson, and singlet-dominated CP-even and CP-odd Higgs bosons, respectively. After considering the XENON-1T constraints, none of them, however, can be fully responsible for the correct abundance. Instead, DM prefered to co-annihilate with the Higgsinos to obtain the density, which corresponds to a correlated parameter space of $2 |\kappa| \simeq \lambda$ with $\lambda \lesssim 0.1$~\cite{Cao:2018rix,cao:2021new-work}. It is worth adding that, with the improvement of the experimental sensitivity in detecting DM, the small upper bound of $\lambda$ will be further reduced.

It is noted that the above conclusions about the Singlino-dominated DM are established in the NMSSM with a $Z_3$ symmetry. In such a framework, to ensure that the neutralino is the LSP,  $|\kappa|$ must be less than $\lambda/2$~\cite{Ellwanger:2009dp}. Consequently, it should be small after considering the DM experiments' substantial limitation on $\lambda$. This conclusion, however, does not hold in the general NMSSM (GNMSSM) due to the emergence of $Z_3$-violating terms. Specifically, the ratio of Singlino and Higgsino masses is no longer $2 |\kappa|/\lambda$,  so $|\kappa|$ may be much larger than $\lambda$ to predict the Singlino-dominated neutralino as the LSP. The Singlino-Higgsino complex system has four independent parameters, which may take $\lambda$, $\mu_{tot}$, $m_{\tilde{\chi}_1^0}$, and $\kappa$. This characteristic contrasts with that of the $Z_3$-NMSSM, which only contains three input parameters, i.e., $\lambda$, $\mu_{tot}$, and any one of $m_{\tilde{\chi}_1^0}$ and $\kappa$. An important application of these differences is that the singlet-dominated particles may form a secluded DM sector~\cite{Pospelov:2007mp}, which has the following salient features.
\begin{itemize}
\item Since the parameter $\kappa$ determines the interactions among the singlet-dominated particles, the Singlino-dominated DM can achieve the correct abundance by the process $\tilde{\chi}_1^0 \tilde{\chi}_1^0 \to h_s A_s$ through adjusting the value of $\kappa$.
\item When the parameter $\lambda$ is small, the secluded sector communicates with the SM sector only through the weak singlet-doublet Higgs mixing. In this case, the interaction between the DM and nucleus is naturally feeble.
\end{itemize}

It is emphasized that $\lambda$ and $\kappa$ may play different roles in the DM physics of the GNMSSM. Similar to the $Z_3$-NMSSM, the parameters $\lambda$, $\mu_{tot}$, and $m_{\tilde{\chi}_1^0}$ mainly affect the coupling between the DM and nucleons, so they have been strongly restricted by DM direct detection (DD) experiments. In contrast, the singlet fields' self-interactions can be entirely responsible for the DM density, where the parameter $\kappa$ plays a crucial role. Owing to this characteristic, the GNMSSM has a broad parameter space consistent with the current DM experimental results. This fact has been neglected in the literature. Considering the dilemma of the $Z_3$-NMSSM in DM physics, the DM physics of the GNMSSM are scrutinized herein. Our results are quite different from those of previous studies on Singlino-dominated DM, which worked in the framework of the $Z_3$-NMSSM ~\cite{Das:2012rr,Ellwanger:2014hia,Ellwanger:2016sur,Ellwanger:2018zxt,Xiang:2016ndq,Cao:2016cnv,
Cao:2018rix,Cao:2019qng,cao:2021new-work,Abdallah:2019znp,Guchait:2020wqn}.

The rest of this paper is organized as follows. First, the key features of the GNMSSM are introduced in Section II. In particular, an economic GNMSSM realization is considered.
The research strategy and numerical results are described in Section III to show the characteristics of DM physics. Finally, conclusions are drawn in Section IV.

\section{\label{theory-section}Theoretical preliminaries}

In this section, the GNMSSM's basics are reviewed. The Singlino-dominated DM's properties will be elucidated in detail by analytic formulae for the case of massive-charge Higgs and gauginos.

\subsection{\label{Section-Model}General NMSSM }

The GNMSSM augments the MSSM by a gauge singlet superfield $\hat{S}$ that does not carry any leptonic or baryonic number. Thus, its Higgs sector contains $\hat{S}$ and two $SU(2)_L$ doublet superfields, $\hat{H}_u=(\hat{H}_u^+,\hat{H}_u^0)$ and $\hat{H}_d=(\hat{H}_d^0,\hat{H}_d^-)$. The general form of its superpotential is~\cite{ Ellwanger:2009dp}
\begin{eqnarray}
W_{\rm GNMSSM}&=& W_{\rm Yukawa} + \lambda \hat{S} \hat{H_u} \cdot \hat{H_d}+\frac{1}{3} \kappa \hat{S}^3+
                \mu \hat{H_u} \cdot \hat{H_d} + \frac{1}{2} \nu \hat{S}^2+\xi \hat{S},
\end{eqnarray}
where $W_{\rm Yukawa}$ is the Yukawa term for quarks and leptons, and $\lambda$ and $\kappa$ are dimensionless coefficients of the trilinear terms invariant under the $Z_3$ symmetry. The terms characterized by the bilinear mass parameters $\mu$ and $\nu$ and the singlet tadpole parameter $\xi$ break the $Z_3$ symmetry.
Without the loss of generality, one can eliminate the linear term in $\hat{S}$ by displacing the field by a specific constant $c$, i.e., $\hat{S} = \hat{S}^\prime + c$,
so that only the bilinear mass terms in $\hat{S}^\prime$ remain to violate the symmetry. As suggested by the studies in~\cite{Lee10090905,Lee11023595,GGRoss11081284,GGRoss12051509}, the $\mu$ and $\nu$'s natural smallness could stem from the breaking of an underlying discrete R symmetry, $Z^R_4$ or $Z^R_8$, at a low-energy scale. Additionally, it is noticeable that a non-minimal coupling $\chi$ of a Higgs bilinear to gravity was introduced to implement a superconformal symmetry in the GNMSSM~\cite{SFerrara10040712,SFerrara10082942}, and this coupling could drive inflation in the early universe~\cite{MEinhorn09122718}. In this case, the extra $\mu$-term is connected with $\chi$ via gravitino mass $m_{3/2}$, i.e., $\mu=\frac{3}{2}m_{3/2}\chi$. It is the only $\rm Z_3$-breaking term in superpotential due to the superconformal symmetry breaking at the Planck scale~\cite{Hollik:2018yek}. In this study, partly motivated by the inflation-inspired model\footnote{It must be emphasized here that, contrary to studies in the inflation-inspired model~\cite{Hollik:2018yek,Hollik:2020plc}, we concern ourselves with the $\mu$ extension's generality irrespective of its origin, and therefore the inflation explanation and its corresponding gravitino DM scenario are not discussed.} and mainly for simplifying our analysis, a specific GNMSSM is investigated in which all dimensional parameters in the superpotential are zero except $\mu$, labeled as $\mu$-extended NMSSM ($\rm \mu NMSSM$) hereafter. The superpotential and corresponding soft supersymmetry-breaking Lagrangian are reduced to the following forms:
\begin{eqnarray}
W_{\rm \mu NMSSM} &=& W_{\rm Yukawa} + (\lambda \hat{S}+\mu) \hat{H_u} \cdot \hat{H_d}+\frac{1}{3} \kappa \hat{S}^3, \nonumber \\
-\mathcal{L}_{soft} & = &\Bigg[A_{\lambda}\lambda S H_u \cdot H_d + \frac{1}{3} A_{\kappa} \kappa S^3+B_\mu \mu H_u\cdot H_d +h.c.\Bigg] \nonumber \\
& & + m^2_{H_u}|H_u|^2 + m^2_{H_d}|H_d|^2 + m^2_{s}|S|^2,
\end{eqnarray}
where $H_u$, $H_d$, and $S$ denote the Higgs superfields' scalar components. Throughout this work, the $B_{\mu}$ term is set to be zero since it plays a minor role in this study and the supersymmetry-breaking mass parameters, $m^2_{H_u}$, $m^2_{H_d}$, and $m^2_{s}$, are fixed by solving the conditional equations to minimize the scalar potential. The Higgs sector is then described by parameters $\lambda$, $\kappa$, $A_\lambda$, $A_\kappa$, $\mu$, and the Higgs fields' vacuum expectation values (vevs)
$\left\langle H_u^0 \right\rangle = v_u/\sqrt{2}$, $\left\langle H_d^0 \right\rangle = v_d/\sqrt{2}$, and $\left\langle S \right\rangle = v_s/\sqrt{2}$
with $v = \sqrt{v_u^2+v_d^2}\simeq 246~\mathrm{GeV}$. The ratio of the vevs $\tan{\beta} \equiv v_u/v_d$ and $\mu_{eff}  \equiv \lambda v_s/\sqrt{2}$ are chosen as input parameters.

Concerning the Higgs sector, it is convenient to work with the field combinations $H_{\rm SM} \equiv \sin \beta  Re[H_u^0] + \cos \beta  Re [H_d^0]$, $H_{\rm NSM} \equiv \cos \beta  Re [H_u^0] - \sin \beta  Re [H_d^0]$, and $A_{\rm NSM} \equiv \cos \beta  Im [H_u^0] - \sin \beta  Im [H_d^0]$~\cite{miller2004,Cao:2012fz}, where $H_{\rm SM}$ represents the Higgs field in the SM with the vev $v/\sqrt{2}$, $H_{\rm NSM}$ denotes the other CP-even doublet Higgs field with a vanishing vev, and $A_{\rm NSM}$ corresponds to the CP-odd Higgs boson in the MSSM. In the bases ($H_{\rm NSM}$, $H_{\rm SM}$, $Re [S]$), the elements of the CP-even Higgs fields' squared mass matrix are formulated as~\cite{Hollik:2020plc}
\begin{eqnarray}
{\cal M}^2_{S, 11}&=& \frac{2 \mu_{eff} (\lambda A_\lambda + \kappa \mu_{eff})}{\lambda \sin 2 \beta} + \frac{1}{2} (2 m_Z^2- \lambda^2v^2)\sin^22\beta, \nonumber \\
{\cal M}^2_{S, 12}&=&-\frac{1}{4}(2 m_Z^2-\lambda^2v^2)\sin4\beta, \nonumber \\
{\cal M}^2_{S, 13}&=&-\frac{1}{\sqrt{2}} ( \lambda A_\lambda + 2 \kappa \mu_{eff}) v \cos 2 \beta, \nonumber \\
{\cal M}^2_{S, 22}&=&m_Z^2\cos^22\beta+ \frac{1}{2} \lambda^2v^2\sin^22\beta,\nonumber  \\
{\cal M}^2_{S, 23}&=& \frac{v}{\sqrt{2}} \left[2 \lambda \mu_{eff} + 2 \lambda \mu - (\lambda A_\lambda + 2 \kappa \mu_{eff}) \sin2\beta \right], \nonumber \\
{\cal M}^2_{S, 33}&=& \frac{\lambda A_\lambda \sin 2 \beta}{4 \mu_{eff}} \lambda v^2   + \frac{\mu_{eff}}{\lambda} (\kappa A_\kappa +  \frac{4 \kappa^2 \mu_{eff}}{\lambda} ) - \frac{\lambda \mu}{2 \mu_{eff}} \lambda v^2, \label{Mass-CP-even-Higgs}
\end{eqnarray}
and in the bases ($A_{\rm NSM}$, $Im [S]$), those for CP-odd Higgs fields are given by
\begin{eqnarray}
{\cal M}^2_{P,11}&=& \frac{2 \mu_{eff} (\lambda A_\lambda + \kappa \mu_{eff})}{\lambda \sin 2 \beta}, \nonumber  \\
{\cal M}^2_{P,22}&=& \frac{(\lambda A_\lambda + 4 \kappa \mu_{eff}) \sin 2 \beta }{4 \mu_{eff}} \lambda v^2  - \frac{3 \mu_{eff}}{\lambda} \kappa A_\kappa - \frac{\lambda \mu}{2 \mu_{eff}} \lambda v^2, \nonumber  \\
{\cal M}^2_{P,12}&=& \frac{v}{\sqrt{2}} ( \lambda A_\lambda - 2 \kappa \mu_{eff}). \label{Mass-CP-odd-Higgs}
\end{eqnarray}
Three CP-even Higgs mass eigenstates $h_i=\{h,H,h_{\rm s}\}$ and two CP-odd Higgs mass eigenstates $a_i=\{A_H, A_{\rm s}\}$ are then obtained by
\begin{eqnarray}
h_i & = & V_{h_i}^{\rm NSM} H_{\rm NSM}+V_{h_i}^{\rm SM} H_{\rm SM}+V_{h_i}^{\rm S} Re[S], \nonumber \\
a_i & = & U_{a_i}^{\rm NSM} A_{\rm NSM}+ U_{a_i}^{\rm S} Im [S],
\end{eqnarray}
where $V$ and $U$ are unitary matrices used to diagonalize $\mathcal{M}^2_S$ and $\mathcal{M}^2_P$, respectively. The model also predicts a pair of charged Higgs bosons, $H^\pm = \cos \beta H_u^\pm + \sin \beta H_d^\pm$, and their masses take the following simple form:
\begin{eqnarray}
  m^2_{H_{\pm}} = \frac{2\mu_{\rm eff}}{\sin 2 \beta} \left(\frac{\kappa}{\lambda}\mu_{\rm eff} + A_{\lambda}\right) + m^2_W -\lambda^2 v^2.
\end{eqnarray}

In this study, the following properties of the Higgs bosons are utilized:
\begin{itemize}
\item $h$ corresponds to the scalar discovered at the LHC. The LHC Higgs data have restricted $\sqrt{\left (V_h^{\rm NSM} \right )^2 + \left ( V_h^{\rm S} \right )^2} \lesssim 0.1$ and $|V_h^{\rm SM}| \sim 1$. Furthermore, from theoretical points of view, its mass may be significantly affected by the interaction $\lambda\,\hat{s}\,\hat{H}_u \cdot \, \hat{H}_d$, the
singlet-doublet Higgs mixing as well as the radiative correction from top/stop loops~\cite{Hall:2011aa,Ellwanger:2011aa,Cao:2012fz}.

\item $H$ and $A_H$ are $H_{\rm NSM}$- and $A_{\rm NSM}$-dominated states. They are approximately degenerate in mass with the
charged Higgs bosons if they are significantly heavier than $v$. The LHC search for extra Higgs bosons, the measured property of $h$, and the indirect constraints from $B$-physics prefer $m_{H}, m_{A_H} \gtrsim 200~{\rm GeV}$~\cite{Baum:2019uzg} when the alignment condition is satisfied~\cite{Carena:2013ooa,Carena:2015moc}. It is worth noting that, although $H$ of several hundreds of GeV may play a significant role in the DM-nucleon scattering discussed below in specific situations~\cite{Huang:2014xua}, its contribution to the scattering is usually far below current DD experimental sensitivities\footnote{The $H$-mediated contribution to the scattering is proportional to $\tan^2 \beta/m_H^4$, and its typical size is $10^{-49}~{\rm cm^2}$ for $\tan \beta \leq 5$ (see, e.g., Figure 6 in~\cite{Baum:2017enm}). With the increase of $\tan \beta$, the lower bound of $m_H$ increases rapidly to satisfy the constraints from the LHC search for extra Higgs bosons (see, e.g., Figure 2 in~\cite{Aad:2020zxo} for the MSSM results.). Therefore, the contribution is difficult to reach $10^{-47}~{\rm cm^2}$.}. Thus, only the case where $H$ and $A_H$ are above $1~{\rm TeV}$ is considered in this paper to simplify the analysis of DM phenomenology. This case is a theoretical hypothesis, and it can be realized by simply setting $A_\lambda$ a significant value, e.g., $A_\lambda = 2~{\rm TeV}$ in the following numerical study. It does not change essentially the conclusions of this paper.

\item $h_s$ and $A_s$ are mainly composed of the singlet field. They may be moderately light (e.g., several tens of GeV) without conflicting with any collider constraints. As will be shown below, they play important roles in DM physics.

\item In the case of massive charged Higgs bosons, the following approximations hold~\cite{Baum:2017enm}:
\begin{eqnarray}
m_{h_s}^2 & \simeq & {\cal M}^2_{S, 33} - \frac{{\cal M}^4_{S, 13}}{{\cal M}^2_{S, 11} - {\cal M}^2_{S, 33}}, \quad m_{A_s}^2 \simeq {\cal M}^2_{P, 22} - \frac{{\cal M}^4_{P, 12}}{{\cal M}^2_{P, 11} - {\cal M}^2_{P, 22}},  \\
\frac{V_{h}^{\rm S}}{V_h^{\rm SM}} & \simeq &  \frac{{\cal M}^2_{S, 23}}{m_h^2 - {\cal M}^2_{S, 33}}, \quad V_{h}^{\rm NSM} \sim 0, \quad V_h^{\rm SM} \simeq \sqrt{1 + \left ( \frac{V_{h}^{\rm S}}{V_h^{\rm SM}} \right )^2}  \sim 1, \nonumber  \\
\frac{V_{h_s}^{\rm SM}}{V_{h_s}^{\rm S}} & \simeq &  \frac{{\cal M}^2_{S, 23}}{m_{h_s}^2 - m_h^2}, \quad V_{h_s}^{\rm NSM} \sim 0, \quad V_{h_s}^{\rm S} \simeq \sqrt{1 + \left ( \frac{V_{h_s}^{\rm SM}}{V_{h_s}^{\rm S}} \right )^2 } \sim 1, \quad U_{A_s}^{\rm S} \simeq 1.  \nonumber
\end{eqnarray}
\end{itemize}
It is emphasized that the mass matrices $\mathcal{M}^2_S$ and $\mathcal{M}^2_P$ differ from their corresponding ones in the $Z_3$-NMSSM by additional $\mu$-induced contributions. These contributions are sometimes vital in this study, e.g., a positively considerable $\mu$ can significantly reduce the mass of $h_s$ and $A_s$ so that they may act as the annihilation product of moderately light DM.

The neutralino sector comprises a Bino field $\tilde{B}^0$, a Wino field $\tilde{W}^0$, Higgsino fields $\tilde{H}_{d}^0$ and $\tilde{H}_u^0$, and a
Singlino field $\tilde{S}^0$. In the bases $\psi^0 = (-i \tilde{B}^0, - i \tilde{W}^0, \tilde{H}_{d}^0, \tilde{H}_{u}^0, \tilde{S}^0)$, the symmetric neutralino mass
matrix is given by~\cite{Ellwanger:2009dp}
\begin{equation}
{\cal M} = \left(
\begin{array}{ccccc}
M_1 & 0 & -m_Z \sin \theta_W \cos \beta & m_Z \sin \theta_W \sin \beta & 0 \\
  & M_2 & m_Z \cos \theta_W \cos \beta & - m_Z \cos \theta_W \sin \beta &0 \\
& & 0 & -\mu_{tot} & - \frac{1}{\sqrt{2}} \lambda v \sin \beta \\
& & & 0 & -\frac{1}{\sqrt{2}} \lambda v \cos \beta \\
& & & & \frac{2 \kappa}{\lambda} \mu_{eff}
\end{array}
\right), \label{eq:MN}
\end{equation}
where $M_1$ and $M_2$ are gaugino soft breaking masses and $\mu_{tot} \equiv \mu_{eff} + \mu$.
It can be diagonalized by a rotation matrix $N$, and, subsequently, the mass eigenstates labeled in mass-ascending order are
\begin{eqnarray}
\tilde{\chi}_i^0 = N_{i1} \psi^0_1 +   N_{i2} \psi^0_2 +   N_{i3} \psi^0_3 +   N_{i4} \psi^0_4 +   N_{i5} \psi^0_5.
\end{eqnarray}
Evidently, $N_{i3}$ and $N_{i4}$ characterize the $\tilde{H}_d^0$ and $\tilde{H}_u^0$ components in $\tilde{\chi}_i^0$, and $N_{i5}$ denotes the Singlino component. We add that the Higgsino mass and Singlino mass in the GNMSSM are determined by $\mu_{tot}$ and $\mu_{eff}$, respectively. As a result, $|2 \kappa/\lambda|$ may be much larger than 1 in obtaining Singlino-dominated DM, which is the starting point of this work.
In addition, there is one more input parameter to describe the Singlino-Higgsino mixing system than in the case of the $Z_3$-NMSSM. These characteristics distinguish the GNMSSM from the $Z_3$-NMSSM, which, as introduced previously, has important implications in DM physics.

In the case of very massive gauginos and $|m_{\tilde{\chi}_1^0}^2 - \mu_{tot}^2 | \gg \lambda^2 v^2$, the following approximations for the Singlino-dominated $\tilde{\chi}_1^0$~\cite{Cheung:2014lqa,Badziak:2015exr,Badziak_2017} are obtained:
\begin{eqnarray}
 \frac{2 \kappa}{\lambda} \mu_{eff}  & \simeq & m_{\tilde{\chi}_1^0} - \frac{1}{2} \frac{\lambda^2 v^2 ( m_{\tilde{\chi}_1^0} - \mu_{tot} \sin 2 \beta )}{m_{\tilde{\chi}_1^0}^2 - \mu_{tot}^2}, \quad N_{11} \sim 0, \quad N_{12} \sim 0, \label{Neutralino-Mixing}  \\
\frac{N_{13}}{N_{15}} &= & \frac{\lambda v}{\sqrt{2} \mu_{\rm tot}} \frac{(m_{\tilde{\chi}_1^0}/\mu_{\rm tot})\sin\beta-\cos\beta} {1-\left(m_{\tilde{\chi}_1^0}/\mu_{\rm tot}\right)^2}, \quad \quad  \frac{N_{14}}{N_{15}} =  \frac{\lambda v}{\sqrt{2} \mu_{\rm tot}} \frac{(m_{\tilde{\chi}_1^0}/\mu_{\rm tot})\cos\beta-\sin\beta} {1-\left(m_{\tilde{\chi}_1^0}/\mu_{\rm tot}\right)^2}, \nonumber \\
N_{15}^2 & \simeq & \left(1+ \frac{N^2_{13}}{N^2_{15}}+\frac{N^2_{14}}{N^2_{15}}\right)^{-1} \nonumber \\
&= & \frac{\left[1-(m_{\tilde{\chi}_1^0}/\mu_{\rm tot})^2\right]^2}{\left[(m_{\tilde{\chi}_1^0}/\mu_{\rm tot})^2
-2(m_{\tilde{\chi}_1^0}/\mu_{\rm tot})\sin2\beta+1 \right]\left(\frac{\lambda v}{\sqrt{2}\mu_{\rm tot}}\right)^2
+\left[1-(m_{\tilde{\chi}_1^0}/\mu_{\rm tot})^2\right]^2}. \nonumber
\end{eqnarray}
These approximations imply the relation
\begin{eqnarray}
\frac{Z_h}{Z_s} &= & \left(\frac{\lambda v}{\sqrt{2} \mu_{\rm tot}}\right)^{\!\!2} \frac{\left(m_{\tilde{\chi}_1^0}/\mu_{\rm tot}\right)^2-2{(m_{\tilde{\chi}_1^0}}/{\mu_{\rm tot}})\sin2\beta+1}
{\left[1-\left({m_{\tilde{\chi}_1^0}}/{\mu_{\rm tot}}\right)^2\right]^2},
\end{eqnarray}
where $Z_h \equiv N_{13}^2+N_{14}^2$ and $Z_s \equiv N_{15}^2$ represent the Higgsino and Singlino fractions in $\tilde{\chi}_1^0$, respectively, and $Z_s > 0.5$ for the Singlino-dominated $\tilde{\chi}_1^0$. They also show that, for fixed $\tan \beta$, the Higgsino compositions in $\tilde{\chi}_1^0$ depend only on $\lambda$, $\mu_{tot}$, and $m_{\tilde{\chi}_1^0}$. In studying the $\tilde{\chi}_1^0$'s properties, it is usually convenient to take the three parameters and $\kappa$ as theoretical inputs.

In this work, the following interactions are crucial:
\begin{eqnarray}
C_{\tilde {\chi}^0_1 \tilde {\chi}^0_1 h_i }  & = &
V_{h_i}^{\rm SM} C_{\tilde {\chi}^0_1 \tilde {\chi}^0_1 H_{\rm SM} }+V_{h_i}^{\rm NSM} C_{\tilde {\chi}^0_1 \tilde {\chi}^0_1 H_{\rm NSM} }+V_{h_i}^{\rm S} C_{\tilde {\chi}^0_1 \tilde {\chi}^0_1 Re[S] } \nonumber \\
& \simeq & \sqrt{2}\lambda \left[ V_{h_i}^{\rm SM} N_{15}\left(N_{13}\sin\beta+N_{14}\cos\beta\right)
+V_{h_i}^{\rm NSM}N_{15}\left(N_{13}\cos\beta- N_{14}\sin\beta\right)\right] \nonumber \\
& & + \sqrt{2} V_{h_i}^{\rm S}
\left(\lambda{N_{13}}{N_{14}}-\kappa N_{15}^2\right)\,,
\label{C-hichi01chi01_S}
\\
C_{\tilde {\chi}^0_1 \tilde {\chi}^0_1 a_i }
& = & U_{a_i}^{\rm NSM}C_{\tilde {\chi}^0_1 \tilde {\chi}^0_1 A_{\rm NSM} }+ U_{a_i}^{\rm S}C_{\tilde {\chi}^0_1 \tilde {\chi}^0_1 Im[S] } \nonumber \\
& \simeq & - \sqrt{2} U_{a_i}^{\rm NSM}\lambda N_{15} ( N_{13}\cos\beta +  N_{14}\sin\beta) + \sqrt{2} U_{a_i}^{\rm S} (\lambda
N_{13}N_{14} - \kappa N_{15}^2), \label{C-aichi01chi01_S} \\
C_{\tilde {\chi}^0_1 \tilde {\chi}^0_1 Z } & = & \frac{m_Z}{v} ( N_{13}^2 - N_{14}^2), \quad \quad C_{\tilde {\chi}^0_1 \tilde {\chi}^0_1 G^0 } \simeq - \sqrt{2} \lambda N_{15} (N_{13} \sin \beta - N_{14}\cos \beta ), \label{C-Zchi01chi01} \\
C_{h_i A_s A_s} &\simeq& \lambda v\, V_{h_i}^{\rm SM} (\lambda + \kappa \sin2\beta) + \lambda \kappa v V_{h_i}^{\rm NSM} \cos2\beta  + \sqrt{2} V_{h_i}^{S}( \sqrt{2} \kappa^2 v_s -\kappa A_\kappa). \label{eq:hiasas}
\end{eqnarray}
These expressions indicate that
\begin{eqnarray}
& & C_{ \tilde{\chi}_1^0 \tilde{\chi}_1^0 h_s} \sim - \sqrt{2}\kappa, \quad C_{\tilde{\chi}_1^0 \tilde{\chi}_1^0 A_s } \sim - \sqrt{2}\kappa, \quad C_{h_s A_s A_s} \sim \sqrt{2} ( \sqrt{2} \kappa^2 v_s -\kappa A_\kappa), \label{Coupling-size} \\
& & C_{h A_s A_s} \sim \lambda v (\lambda + \kappa \sin2\beta), \quad C_{H A_s A_s} \sim \lambda \kappa v, \quad  |C_{h_s A_s A_s}| \gg |C_{h A_s A_s}|, |C_{H A_s A_s}|. \nonumber
\end{eqnarray}
and the other interactions are suppressed.

\subsection{DM Relic Density }

In the GNMSSM, Singlino-dominated DM could achieve the measured abundance by the processes $\tilde{\chi}_1^0 \tilde{\chi}_1^0 \to h_s A_s, t \bar{t}$, or its co-annihilation with one or more slightly heavier states, such as the next-to-lightest neutralino $\tilde{\chi}_2^0$, the lightest chargino $\tilde{\chi}_1^\pm$, or sleptons.
Under a non-relativistic approximation, the thermally averaged cross-section of DM pair annihilation can be expanded as~\cite{Jungman1995Supersymmetric,griest1991three}
\begin{equation}
\label{thermalcs}
\left\langle \sigma_Av \right\rangle = a + b\left\langle v^2 \right\rangle + \mathcal{O}(\left\langle v^4 \right\rangle) \approx a + 6 \frac{b}{x}
\end{equation}
where $a$ and $b$ correspond to the s- and p-wave contributions of the process, respectively, and $x\equiv m_{\tilde{\chi}_1^0}/T$ with $T$ denoting the temperature in the early Universe.
After integrating the density function from $x_F=m/T_F$ at freeze-out temperature to infinity, the present relic density can be expressed as~\cite{Baum:2017enm}
\begin{eqnarray}\label{density}
\Omega h^2 = 0.12\left(\frac{80}{g_*}\right)^{1/2}\left(\frac{x_F}{25}\right) \left( \frac{2.3\times 10^{-26} \mathrm{cm^3/s}}{\langle \sigma v\rangle_{x_F}}\right)\;, \label{density}
\end{eqnarray}
where $g_*\sim 80$ is the number of effectively relativistic degrees of freedom, and $x_F\sim 25$ is obtained by solving the freeze-out equation in~\cite{griest1991three}.

The key features of these annihilations are shown in the following.

1. $\tilde{\chi}_1^0 \tilde{\chi}_1^0 \to h_s A_s$

This process proceeds through the $s$-channel exchange of $Z$ and CP-odd Higgs bosons, and the $t$-channel exchange of neutralinos. Consequently, $\langle \sigma v \rangle_{x_F}^{h_s A_s} $ is given by~\cite{Baum:2017enm,griest1991three}
\begin{eqnarray} \label{eq:sigvPhiPhi}
\left\langle \sigma v \right\rangle_{x_F}^{h_s A_s} \simeq && \frac{1}{64 \pi m_{\tilde{\chi}_1^0}^2} \left\{ \left[1-\frac{\left(m_{h_s} + m_{A_s}\right)^2}{4 m_{\tilde{\chi}_1^0}^2}\right] \left[1-\frac{\left(m_{h_s} - m_{A_s}\right)^2}{4 m_{\tilde{\chi}_1^0}^2}\right] \right\}^{1/2} | {\cal{A}}_s + {\cal{A}}_t |^2, \nonumber
\end{eqnarray}
where the $s$- and $t$-channel contributions are approximated by
\begin{eqnarray}
{\cal{A}}_s &\simeq & \frac{-2 m_{\tilde{\chi}_1^0} C_{\tilde {\chi}^0_1 \tilde {\chi}^0_1 A_s }  C_{h_s A_s A_s}}{m_{A_s}^2 - 4  m_{\tilde{\chi}_1^0}^2}, \nonumber \\
{\cal{A}}_t &\simeq & - 2 C_{\tilde{\chi}_1^0\tilde{\chi}_1^0 h_s} \, C_{\tilde{\chi}_1^0\tilde{\chi}_1^0 A_s} \left[ 1 + \frac{ 2 m_{A_s}^2}{ 4 m_{\tilde{\chi}_1^0}^2 - \left(m_{h_s}^2 + m_{A_s}^2\right) } \right],
\end{eqnarray}
respectively, if there are no resonant contributions. With the formulae from Eqs.(\ref{C-hichi01chi01_S})--(\ref{Coupling-size}), one can learn that,
once the involved particles' masses are fixed, the density is mainly determined by the parameter $\kappa$. The authors of Ref.~\cite{Baum:2017enm} estimated the value of $\kappa$ to predict the density, and concluded that
\begin{eqnarray}
|\kappa| \sim 0.15 \times \left ( \frac{m_{\tilde{\chi}_1^0}}{300~{\rm GeV}} \right )^{1/2}
\end{eqnarray}
in the case of $|{\cal{A}}_t| \gg |{\cal{A}}_s|$.

2. $\tilde{\chi}_1^0 \tilde{\chi}_1^0 \to t \bar{t}$

This annihilation is mediated by the s-channel exchange of $Z$ and Higgs bosons. Owing to the largeness of top quark mass, its amplitude is mainly contributed by the longitudinal polarization of the $Z$ boson when the mediators of the Higgs bosons are far off-shell. As a result, $\langle \sigma v \rangle_{x_F}^{t \bar{t}}$ is approximated by~\cite{Baum:2017enm}
\begin{eqnarray} \label{eq:gxxGforOmega}
\langle \sigma v \rangle_{x_F}^{t\bar{t}} \sim 2 \times 10^{-26}\,\frac{{\rm cm}^3}{{\rm s}} \left( \frac{\left|C_{\tilde{\chi}_1^0 \tilde{\chi}_1^0  G^0}\right|}{0.1} \right)^2 \left( \frac{m_{\tilde{\chi}_1^0} }{300\,{\rm GeV}}\right)^{-2},
\end{eqnarray}
where
\begin{eqnarray} \label{eq:xsxsG}
C_{\tilde{\chi}_1^0 \tilde{\chi}_1^0 G^0} &\simeq & - \sqrt{2} \lambda N_{15} \left( N_{13} s_\beta - N_{14} c_\beta \right)  \simeq \frac{\lambda^2 v}{\mu_{tot}} \frac{Z_s(m_{\tilde{\chi}_1^0}/\mu_{tot})\cos2\beta}{1-(m_{\tilde{\chi}_1^0}/\mu_{tot})^2}.
\end{eqnarray}
These formulae reflect that a moderately large $\lambda$ is needed to account for the density. As will be shown below, this situation usually leads to sizable DM-nucleon scattering rates.

3. Co-annihilation with Higgsino-dominated electroweakinos

This annihilation affects the abundance when the mass splitting between $\tilde{\chi}_1^0$ and the Higgsino-dominated electroweakinos is less than approximately $10\%$~\cite{griest1991three,Coannihilation}.
In this case, $\sigma$ in Eq.(\ref{density}) should be replaced by $\sigma_{eff}$ given by
\begin{eqnarray}
    \sigma_{eff} &&= \sum^N_{a,b}\sigma_{ab}\frac{g_ag_b}{g^2_{eff}}(1+\Delta_a)^{3/2}(1+\Delta_b)^{3/2}\times exp[-x(\Delta_a+\Delta_b)]. \label{Co-annihilation experession}
\end{eqnarray}
In this formula, $\sigma_{ab}$ is the cross-section of the annihilation $a b \to X Y$ with $a, b = \tilde{\chi}_1^0, \tilde{\chi}_2^0, \tilde{\chi}_1^\pm, \cdots$ and $X, Y$ denoting any possible final states, $\Delta_i \equiv (m_i - m_{\tilde{\chi}_1^0})/m_{\tilde{\chi}_1^0}$ ($i=a, b$) represents the mass splitting between $\tilde{\chi}_1^0$ and the initial state $i$, and $x \equiv m_{\tilde{\chi}_1^0}/T$. $g_i$ denotes the internal degrees of freedom for the initial state $i$, which is 2 for a Majorana fermion and 4 for a Dirac fermion, and $g_{eff}$ is defined by
\begin{eqnarray}
g_{eff} \equiv   \sum^N_{a} g_a(1+\Delta_a)^{3/2}exp(-x\Delta_a).
\end{eqnarray}
It is evident that $\sigma_{eff}$ decreases monotonously as the co-annihilating particles depart from $\tilde{\chi}_1^0$ in mass.

Finally, the annihilations $\tilde{\chi}_1^0 \tilde{\chi}_1^0 \to h A_s, h_s h_s$ are briefly discussed. The former process proceeds in a similar way as $\tilde{\chi}_1^0 \tilde{\chi}_1^0 \to h_s A_s$.  From the expressions of $ C_{\tilde{\chi}_1^0 \tilde{\chi}_1^0 h}$ and $C_{h A_s A_s}$ and the fact that $C_{\tilde{\chi}_1^0 \tilde{\chi}_1^0 h_s} \gg C_{\tilde{\chi}_1^0 \tilde{\chi}_1^0 h}$ and $C_{h_s A_s A_s} \gg C_{h A_s A_s}$ in most cases, one can conclude that the abundance is significantly affected only when $\lambda$ and $V_h^S$ are tremendously large. This condition, however, has been tightly limited by the XENON-1T and LHC experiments.
Process $\tilde{\chi}_1^0 \tilde{\chi}_1^0 \to h_s h_s$ proceeds by the s-channel exchange of the CP-even Higgs bosons and the t-channel exchange of the neutralinos. It is a p-wave process~\cite{Griest:1989zh}, so its effect on the abundance becomes crucial only when $\kappa$ is large and $\tilde{\chi}_1^0 \tilde{\chi}_1^0 \to h_s A_s$ is phase-space-suppressed. This situation happens only in some corners of the theory's parameter space.

\subsection{DM-nucleon scattering cross-sections}

In the limit of very massive squarks, the SD scattering of $\tilde{\chi}_1^0$ with nucleons is mediated by a $Z$ boson. The scattering cross-section is then given by~\cite{Badziak:2015exr,Badziak_2017}
\begin{eqnarray}
\label{eq:sigSD}
\sigma_{\tilde{\chi}_1^0-N}^{\rm SD} & \simeq & C_N \times \left(\frac{N_{13}^2-N_{14}^2}{0.1}\right)^2,
\end{eqnarray}
where $C_p \simeq 4 \times 10^{-4}\rm~pb$ for protons and $C_n \simeq 3.1 \times 10^{-4}\rm~pb$ for neutrons, and
\begin{eqnarray}
\label{eq:N13N14diff}
N_{13}^2-N_{14}^2 =  \big( \frac{\lambda v}{\sqrt{2}\mu_{tot}} \big)^2
\, \frac{Z_s\cos2\beta} {1-(m_{\tilde{\chi}_1^0}/\mu_{tot})^2},
\end{eqnarray}
from the formulae in Eq.(\ref{Neutralino-Mixing}). These expressions indicate that the SD cross-section is proportional to $(\lambda v/\mu_{tot})^4$ and it increases as $m_{\tilde{\chi}_1^0}$ approaches from below to $\mu_{tot}$.

The SI scattering is mainly contributed by the $t$-channel exchange of the CP-even Higgs bosons.
Correspondingly, the cross-section is given by~\cite{badziak2016blind,pierce2013neutralino}
\begin{eqnarray}
\label{SI-N}
\sigma^{SI}_{\tilde {\chi}^0_1-{N}} = \frac{4 \mu_r^2}{\pi} |f^{(N)}|^2 , \label{SI-cross}
\end{eqnarray}
where $N=p, n$ denotes proton and neutron, respectively, and $\mu_r$ is the reduced mass of $\tilde{\chi}_1^0$ and nucleon given by $\mu_r \equiv m_N m_{\tilde {\chi}^0_1} /(m_N+m_{\tilde {\chi}^0_1})$. The effective couplings $f^{(N)}$ are
\begin{eqnarray}
\label{SI-fN}
f^{(N)} =  \sum_{h_i=\{h,H,h_{\rm s}\}} \!f^{(N)}_{h_i} = \sum_{h_i=\{h,H,h_{\rm s}\}}\! \frac{C_{ \tilde {\chi}^0_1 \tilde {\chi}^0_1 h_i} C_{N N h_i }}{2m^2_{h_i} },
\end{eqnarray}
where $C_{ N N h_i}$ represents the $h_i$-nucleon coupling, which is given by
\begin{eqnarray}
\label{C-hNN_S}
C_{NN h_i} = -\frac{m_N}{v}
\left[ F^{(N)}_d \left( V_{h_i}^{\rm SM}- \tan\beta V_{h_i}^{\rm NSM}\right)+F^{(N)}_u \left(V_{h_i}^{\rm SM}+\frac{1}{\tan\beta} V_{h_i}^{\rm NSM} \right)
\right]\,,
\end{eqnarray}
with $F^{(N)}_d \equiv f^{(N)}_d+f^{(N)}_s+\frac{2}{27}f^{(N)}_G$ and $F^{(N)}_u \equiv f^{(N)}_u+\frac{4}{27}f^{(N)}_G$. The nucleon form factors $f^{(N)}_q =m_N^{-1}\left<N|m_qq\bar{q}|N\right>$ for $q=u,d,s$ represent the normalized light quark contribution to nucleon mass, and $f^{(N)}_G=1-\sum_{q=u,d,s}f^{(N)}_q$ denotes other heavy quarks' mass fraction in the nucleon~\cite{Drees1993,Drees1992}. For the default settings of the package micrOMEGAs about $f_q^{N}$~\cite{Belanger_2009,Alarcon:2011zs,Alarcon:2012nr},
$F_u^{p} \simeq F_u^n \simeq 0.15$ and $F_d^{p} \simeq F_d^n \simeq 0.13$. Consequently, one can conclude that $\sigma^{SI}_{\tilde {\chi}^0_1-p} \simeq \sigma^{SI}_{\tilde {\chi}^0_1-n}$.

In the case of very massive charged Higgs bosons, $V_h^{\rm NSM} \sim 0$, $V_{h_s}^{\rm NSM} \sim 0$, and the $H$-mediated contribution can be safely neglected since it is further suppressed by a factor $1/m_H^4$. Consequently, the SI cross-section is given by~\cite{cao:2021new-work}
\begin{eqnarray}
\sigma^{\rm SI}_{\tilde{\chi}_1^0-N} & \simeq & 5 \times 10^{-45}\,{\rm cm^2}~\left(\frac{\cal{A}}{0.1}\right)^2,
\end{eqnarray}
where
\begin{eqnarray}
{\cal{A}} = \left( \frac{125 \rm GeV}{m_{h}}\right)^2 V_h^{\rm SM} C_{ \tilde {\chi}^0_1 \tilde {\chi}^0_1 h} +  \left( \frac{125 \rm GeV}{m_{h_s}} \right)^2  V_{h_s}^{\rm SM} C_{ \tilde {\chi}^0_1 \tilde {\chi}^0_1 h_{s}}~.
\end{eqnarray}
Furthermore, using the formulae from Eqs.(\ref{Neutralino-Mixing})--(\ref{eq:hiasas}), one may conclude that
\begin{eqnarray}
C_{\tilde {\chi}^0_1 \tilde {\chi}^0_1 h}
& \simeq & \frac{\mu_{tot}}{ v}\,\big( \frac{\lambda v}{\mu_{tot}} \big)^2\, \frac { Z_s V_{h}^{\rm SM}(m_{\tilde{\chi}_1^0}/\mu_{tot} -\sin 2 \beta)}{1-(m_{\tilde{\chi}_1^0}/\mu_{tot})^2}  + \frac{\lambda}{2 \sqrt{2}} \big( \frac{\lambda v}{\mu_{tot}} \big)^2 \frac{Z_s V_{h}^{\rm S} \sin2\beta}{\big[ 1-(m_{\tilde{\chi}_1^0}/\mu_{tot})^2 \big]}\nonumber \\
& & -\sqrt{2}\kappa Z_s V_h^{\rm S} \left[1+ \big( \frac{\lambda v}{\sqrt{2}\mu_{tot}} \big)^2\frac{1}{1-(m_{\tilde{\chi}_1^0}/\mu_{tot})^2} \frac{\mu_{eff}}{\mu_{tot}} \right],
\label{C_xsxshsm} \\
C_{\tilde {\chi}^0_1 \tilde {\chi}^0_1 h_{s}} & \simeq &
\frac{\mu_{tot}}{ v}\,\big( \frac{\lambda v}{\mu_{tot}} \big)^2\, \frac { Z_s V_{h_{s}}^{\rm SM}(m_{\tilde{\chi}_1^0}/\mu_{tot} -\sin 2 \beta)}{1-(m_{\tilde{\chi}_1^0}/\mu_{tot})^2}  + \frac{\lambda}{2 \sqrt{2}} \big( \frac{\lambda v}{\mu_{tot}} \big)^2 \frac{Z_s V_{h_{s}}^{\rm S} \sin2\beta}{\big[ 1-(m_{\tilde{\chi}_1^0}/\mu_{tot})^2 \big]}\nonumber \\
& & -\sqrt{2}\kappa Z_s V_{h_{s}}^{\rm S} \left[1+ \big( \frac{\lambda v}{\sqrt{2} \mu_{tot}} \big)^2\frac{1}{1-(m_{\tilde{\chi}_1^0}/\mu_{tot})^2} \frac{\mu_{eff}}{\mu_{tot}}  \right].
\label{C_xsxshs}
\end{eqnarray}
These formulae indicate that, if the LHC-discovered scalar is pure SM-like, i.e., $V_h^{\rm SM}=1$, $V_h^{\rm S}=0$, and $V_{h_s}^{\rm SM}=0$, the expression of ${\cal{A}}$
is simplified as
\begin{eqnarray}
{\cal{A}} = \left( \frac{125 \rm GeV}{m_{h}}\right)^2 \frac{\mu_{tot}}{v}\,\big( \frac{\lambda v}{\mu_{tot}} \big)^2\, \frac { Z_s (m_{\tilde{\chi}_1^0}/\mu_{tot} -\sin 2 \beta)}{1-(m_{\tilde{\chi}_1^0}/\mu_{tot})^2}.
\end{eqnarray}
To suppress the SI cross-section, either $\lambda$ is small or $m_{\tilde{\chi}_1^0}/\mu_{\rm tot} \simeq \sin2\beta$, which is called the blind-spot condition of the NMSSM~\cite{Badziak:2015exr, badziak2016blind}. However, if $h_s$ contains a sizable SM Higgs field component, e.g., $V_{h_s}^{\rm SM} > 0.1$, the $h_s$-mediated contribution may be crucial when $h_s$ is much lighter than $h$. This may drastically cancel the $h$-mediated contribution. In either case, the SI cross-section is usually dominated by the $h$-mediated contribution, and it is sensitive to the parameters $\lambda$ and $\mu_{tot}$. Note that this feature still holds when the $H$-mediated contribution is not negligibly small. Therefore, the DM direct detection experiments can tightly limit these two parameters. Finally, it is noticeable that $\kappa$ may also affect the cross-section significantly, but it occurs only when $\kappa \gg \lambda$ and $V_h^{\rm S}$ is sizable. This situation is different from that of the SD cross-section, which is insensitive to $\kappa$ in all cases.

\section{Numerical Results}

In this section, the characteristics of the Singlino-dominated DM are demonstrated by numerical results. A likelihood function is constructed for the nest sampling algorithm adopted in this work to guide sophisticated scans over the $\mu$NMSSM's parameter space~\cite{Feroz:2008xx,Feroz:2013hea}. The scanned samples are studied by statistical quantities such as the posterior probability density function (PDF) in Bayesian statistics and the profile likelihood (PL) in frequency statistics. The former reflects the preference of the obtained samples to the input parameters, and the latter, however, represents the theory's capability to explain experimental data.
These statistical quantities have been introduced concisely in~\cite{Fowlie:2016hew} and are applied to scrutinize the phenomenology of the NMSSM's seesaw extensions~\cite{Cao:2019qng,cao2019bayesian,Cao:2019aam}.

\subsection{Research Strategy}

\begin{table}
\centering
\begin{tabular}{|c|c|c|c|c|c|}
\hline Parameter & Value & Parameter & Value & Parameter & Value \\
\hline
$A_{\lambda}$ & $2 \mathrm{TeV}$ & $m_{\tilde{q}}$ & $2 \mathrm{TeV}$ & $m_{\tilde{l}}$ & $2 \mathrm{TeV}$ \\
$M_{1}$ & $3 \mathrm{TeV}$ & $M_{2}$ & $3 \mathrm{TeV}$ & $M_{3}$ & $3 \mathrm{TeV}$ \\
\hline
\end{tabular}
\caption{Setting of different fields' soft-breaking parameters. $A_\lambda$, $m_{\tilde{q}}$, $m_{\tilde{l}}$, and $M_i$ (i=1,2,3) are for Higgs, squarks, sleptons, and different gaugino fields, respectively. \label{tab:tab1}}
\end{table}

The following likelihood function is taken:
\begin{eqnarray}
\mathcal{L}= \mathcal{L}_{\Omega h^2} \times \mathcal{L}_{DD} \times \mathcal{L}_{IDD} \times \mathcal{L}_{Higgs} \times \mathcal{L}_{B}  \label{Likelihood}
\end{eqnarray}
where $\mathcal{L}_{\Omega h^2} $ is a Gaussian likelihood function for the measured DM abundance, $\mathcal{L}_{DD}$ encodes the upper bounds on the SI and SD cross-sections from the XENON-1T experiments~\cite{Aprile:2018dbl,Aprile:2019dbj}, and $\mathcal{L}_{IDD}$ incorporates the DM indirect search information of the Fermi-LAT experiment~\cite{Ackermann:2015zua,FermiLat}.
$\mathcal{L}_{Higgs}$ includes a fit of the $h$'s property to corresponding data of the LHC by the code \textsf{HiggsSignal-2.2.3}~\cite{HiggsSignal}. It also includes the constraints from the direct search for extra Higgs bosons at the LEP, Tevatron, and LHC, which are implemented by the code \textsf{HiggsBounds-5.3.2}~\cite{HiggsBounds}.
Finally, $\mathcal{L}_{B}$ takes a Gaussian form to describe the measured branching ratios of $B \to X_s \gamma$ and $B^0_s \to \mu^+ \mu^-$~\cite{Tanabashi:2018oca}. The concrete expression of $\mathcal{L}$ was given in our previous works~\cite{cao2019bayesian,Cao:2019aam}.

The following parameter space is scanned:
\begin{eqnarray}
&& 0<\lambda \leq 0.70,\quad |\kappa| \leq 0.70, \quad 1 \leq {\rm tan} \beta \leq 60, \quad |A_t| \leq 5~{\rm TeV},  \label{scan_range} \\
&& 0 < \mu_{eff} \leq 500 ~{\rm GeV}, \quad 100~{\rm GeV} \leq |\mu_{tot}| \leq 500 ~{\rm GeV}, \quad  |A_{\kappa}| \leq 1000 ~{\rm GeV},  \nonumber
\end{eqnarray}
by assuming the input parameters are flat distributed. Upper bounds are imposed on $\lambda$ and $\kappa$ to keep the theory perturbative up to the Grand Unification scale. $\mu_{eff}$, $\mu_{tot}$, and $A_\kappa$ are restricted within relatively narrow ranges to naturally break the electroweak symmetry, and the parameter $A_t$ is varied to include the critical radiative correction of the top-stop loops to Higgs mass. In addition, a lower bound is set on $|\mu_{tot}|$ by the LEP search for electroweakinos~\cite{Tanabashi:2018oca}. For the other less crucial parameters, their values are fixed as in Table~\ref{tab:tab1}.

The package \textsf{SARAH-4.14.3}~\cite{Staub08060538,Staub12070906,Staub13097223,Staub2015exploring} was utilized to build the $\mu$NMSSM's model file, and the codes  \textsf{SPheno-4.0.3}~\cite{Porod:2003um,Porod_2012}  and \textsf{FlavorKit}~\cite{Porod:2014xia} were used to generate the particle spectrum and calculate low-energy
flavor observables, respectively. The Singlino-dominated neutralino is assumed to be the sole DM candidate in the Universe and the observables in DM physics are computed with the package \textsf{MicrOMEGAs 5.0.4}~\cite{Belanger:2001fz,Belanger:2005kh,Belanger:2006is,belanger2010micromegas,Belanger:2013oya,barducci2017collider}.  We consider $h \equiv h_1$ and $h \equiv h_2$ scenarios, where $h_1$ and $h_2$ denote the lightest and the next-to-lightest Higgs bosons.
Details of each obtained sample are reported for further analysis. For example, all processes that affect the abundance are listed and their fractions to the total DM
annihilation cross-section at the freeze-out temperature are presented. This information helps identify the dominant contribution to the abundance.

\begin{table}[]
\begin{center}
\begin{tabular}{|c|c|c|c|c|}
\hline
                    & \multicolumn{2}{c|}{$h \equiv h_1$ scenario} & \multicolumn{2}{c|}{$h \equiv h_2$ scenario} \\ \hline
Parameters    & 1 $ \sigma$         & 2 $ \sigma$         & 1 $\sigma$         & 2 $\sigma$         \\ \hline
$\lambda$     & 0.04 $\sim$ 0.24   & 0.02 $\sim$ 0.41  & 0.05 $\sim$ 0.18  & 0.03 $\sim$ 0.31  \\ \hline
$\kappa$      & -0.25 $\sim$ 0.31 & -0.31 $\sim$ 0.41 & -0.17 $\sim$ 0.22 & -0.30 $\sim$ 0.30 \\ \hline
$\tan{\beta}$ & 5.9 $\sim$ 23.3  & 4.7 $\sim$ 49.3  & 5.7 $\sim$ 12.1  & 5.1 $\sim$ 20.2  \\ \hline
$\mu$         & 245 $\sim$ 420  & 163 $\sim$ 467  & 150 $\sim$ 319  & 51 $\sim$ 408  \\ \hline
$\mu_{eff}$         & 29 $\sim$ 118  & 14 $\sim$ 253  & 33 $\sim$ 148  & 18 $\sim$ 298  \\ \hline
$\mu_{tot}$ & 325 $\sim$ 475         & 255 $\sim$ 495        & 225 $\sim$ 425         & 136 $\sim$ 475        \\ \hline
$A_{\kappa}$        & -375 $\sim$ 128        & -576 $\sim$ 397       & -652 $\sim$ 339        & -776 $\sim$ 534       \\ \hline
$A_t$         & 3148 $\sim$ 4248     & 2849 $\sim$ 4448    & 2651 $\sim$ 4150    & -3845 $\sim$ 4550   \\ \hline
$m_{\tilde{\chi}_1^0}$ & 240 $\sim $ 407 & 182 $\sim $ 466 & 181 $\sim$ 346 & 118 $ \sim $ 404 \\ \hline
\end{tabular}
\caption{One-dimensional 1$\sigma$ and 2$\sigma$ credible regions for input parameters in $h \equiv h_1$ and $h \equiv h_2$ scenarios, where $h$, $h_1$, and $h_2$ denote the SM-like Higgs boson and the lightest and next-to-lightest CP-even Higgs boson, respectively. Dimensional parameters are in units of {\rm GeV}.}\label{tab:tab2}
\end{center}
\end{table}

\begin{table}[]
\begin{center}
\begin{tabular}{|c|c|c|c|}
\hline
\multicolumn{4}{|c|}{$h \equiv h_1$ scenario: lnZ = -65.79 $\pm$ 0.046 } \\ \hline
$ \rm \quad  \tilde{\chi}_1^0 \tilde{\chi}_1^0 \; \to \; h_s \; A_s$ &
  $\rm \quad  \tilde{\chi}_1^0 \tilde{\chi}_1^0 \; \to \; t \; \bar{t}$ &
  $\rm \quad \; \, \tilde{\chi}_1^0 \tilde{\chi}_1^0 \; \to \; h_s \; h_s$ &
  Co-annihilation \\ \hline
$88\%$              & $8\%$              & $3\%$              & $0.7\%$             \\ \hline
\multicolumn{4}{|c|}{$h \equiv h_2$ scenario: lnZ = -68.23 $\pm$ 0.051} \\ \hline
$ \rm \quad  \tilde{\chi}_1^0 \tilde{\chi}_1^0 \; \to \; h_s \; A_s$ &
  $\rm \quad  \tilde{\chi}_1^0 \tilde{\chi}_1^0 \; \to \; t \; \bar{t}$ &
  Co-annihilation & $h$-funnel
   \\ \hline
$76\%$              & $12\%$             & $11.6\%$             &  $0.3\%$     \\ \hline
\end{tabular}
\caption{Dominant annihilation channels and their normalized posterior probabilities for $h \equiv h_1$ and $h \equiv h_2$ scenarios.
 In obtaining the values in this table, each sample's most critical channel for the abundance was identified and sequentially used to classify the samples. The posterior probability densities of the same type of samples were then summed.
}\label{tab:tab3}
\end{center}
\end{table}

\subsection{Posterior probability density}

In Bayesian statistics, the one-dimensional marginal posterior PDF $P(\theta)$ with $\theta$ being an input parameter is defined by integrating the posterior PDF over the other input parameters. This decides the one-dimensional $1\sigma$ ($2\sigma$) credible region by requiring that the region's posterior probability accounts for $68\%$ ($95\%$) of the total probability~\cite{Fowlie:2016hew}. In this study, each sample's posterior probability density is calculated by the nest sampling algorithm and the $1\sigma$ and $2 \sigma$ credible regions for the input parameters are determined. The results are presented in Table~\ref{tab:tab2}. Compared with the scanned region, one can learn that some parameters' favored range is reduced significantly. For example, $\lambda$ and $|\kappa|$ prefer the range $\lambda \lesssim 0.4 $ and $|\kappa| \lesssim 0.3 $ because, as mentioned above, it is more easily consistent with various DM measurements.

In Table~\ref{tab:tab3}, dominant annihilation channels for the obtained samples and their posterior probability are listed. This table shows that $\tilde{\chi}_1^0 \tilde{\chi}_1^0 \to h_s A_s$ is the dominant channel for most cases, and $\tilde{\chi}_1^0 \tilde{\chi}_1^0 \to t \bar{t}$ is another popular channel. In contrast, the co-annihilation is vital only in rare cases in the $h\equiv h_1$ scenario. The characteristics of these channels were introduced in the preceding section. In the following, only several new features after
elaborated analysis of the samples are listed.
\begin{itemize}
\item Any of $\tilde{\chi}_1^0 \tilde{\chi}_1^0 \to h_s A_s$, the co-annihilation, and the $h$-funnel can be fully responsible for the DM abundance. Compared with the first mechanism for the abundance, the latter two need a stricter condition, i.e., $|m_{\tilde{\chi}_1^0}| \simeq |\mu_{tot}|$ and $m_{\tilde{\chi}_1^0} \simeq m_h/2$, respectively. Therefore, their Bayesian evidence is relatively small\footnote{Bayesian evidence is usually defined by $Z(D|M) \equiv \int{P(D|O(M,\Theta)) P(\Theta|M) \prod d \Theta_i}$, where $P(\Theta|M)$ is the prior PDF of input parameters $\Theta = (\Theta_1,\Theta_2,\cdots)$ in model $M$ and $P(D|O(M, \Theta))\equiv \mathcal{L}(\Theta)$ is the likelihood function for observables $O$, which considers both theoretical predictions, $O(M, \Theta)$, and experimental data, $D$. Computationally, the evidence represents an average likelihood
     in the surveyed parameter space. It depends on  the priors of the model's parameters. For different scenarios in a theory, the larger $Z$ is, the more readily it is for the corresponding scenario to agree with the data. The details of Bayesian evidence and its applications were presented in, e.g.,~\cite{Bayes}.}.
\item Owing to the tight constraints of the XENON-1T experiments on $\lambda$, $\tilde{\chi}_1^0 \tilde{\chi}_1^0 \to t \bar{t}$ contributes to the total annihilation cross-section at the freeze-out temperature by less than $50\%$. This fact implies that, although $\tilde{\chi}_1^0 \tilde{\chi}_1^0 \to t \bar{t}$ is the most considerable contribution in some cases, it must be aided by other processes to achieve the correct abundance.
\item Owing to its p-wave nature, $\tilde{\chi}_1^0 \tilde{\chi}_1^0 \to h_s h_s$ contributes to the abundance by, at most, $40\%$. This situation is quite similar to that of $\tilde{\chi}_1^0 \tilde{\chi}_1^0 \to t \bar{t}$. Furthermore, since $\tilde{\chi}_1^0 \tilde{\chi}_1^0 \to h_s h_s$ dominates over $\tilde{\chi}_1^0 \tilde{\chi}_1^0 \to h_s A_s$ only when the strict conditions of $ m_{\tilde{\chi}_1^0} > m_{h_s}$, $2 |m_{\tilde{\chi}_1^0}| \lesssim (m_{h_s} + m_{A_s})$ and $|\kappa| \gtrsim 0.3$ are satisfied, its evidence tends to be small.
\end{itemize}

The next object of focus is the Bayesian evidence for different scenarios. Our calculation shows that $\ln Z = - 65.79 \pm 0.046$ for the $h \equiv h_1$ scenario and $\ln Z = - 68.21 \pm 0.051$ for the $h \equiv h_2$ scenario. Since Jeffreys' scale defined by $ \delta \ln Z \equiv \ln (Z_{h_1}/ Z_{h_2})$~\cite{Jeffreys} is $2.42$, it is concluded that the considered experiments prefer the $h \equiv h_1 $ scenario slightly to the $h \equiv h_2$ scenario~\footnote{Given two scenarios to be compared together, Jeffreys' scale provides a calibrated spectrum of significance for the relative strength of the scenarios~\cite{Jeffreys}. For the application of the Jeffreys' scale in particle physics, see, for e.g.,~\cite{Feroz:2008wr}. }. One reason for this conclusion is that the latter scenario needs tuning of its parameter space to explain the $125{\rm GeV}$ Higgs data~\cite{Cao:2012fz}. Also compared in the present paper is the Bayesian evidence of the $\mu \neq 0$ and $\mu = 0$ cases (note that the $\mu = 0$ case corresponds to the $Z_3$-NMSSM), and it is found that the Jeffrey's-scale result is $8.05$. This result implies that the experiments strongly favor the $\mu \neq 0$ case because it is readily consistent with the DM experiments. Realizing that the evidence relies on the prior probability assumptions, it is recalculated by assuming that $\lambda$ and $\kappa$ are log-distributed and the other input parameters are flat-distributed. It is found that Jeffrey's scale changes only slightly after recalculation.

Finally, the experimentally favored parameter space of the $\mu$NMSSM is compared with that of the $Z_3$-NMSSM, which was obtained in our recent work~\cite{cao:2021new-work}. It is found that they are quite different. The reason is, as mentioned before, that the additional parameter $\mu$ contributes to the Higgs and neutralino mass matrices, and, consequently, the DM physics differs significantly.

\begin{figure*}[!t]
		\centering
		\resizebox{0.98\textwidth}{!}{
        \includegraphics{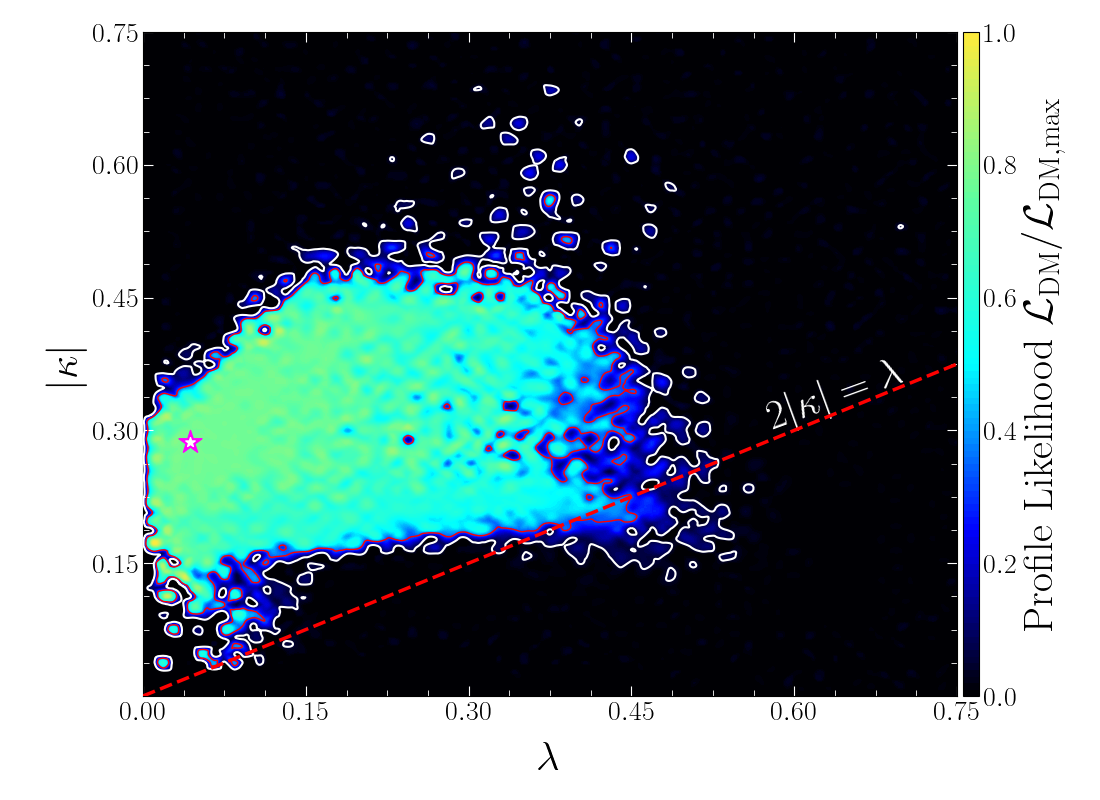}
        \includegraphics{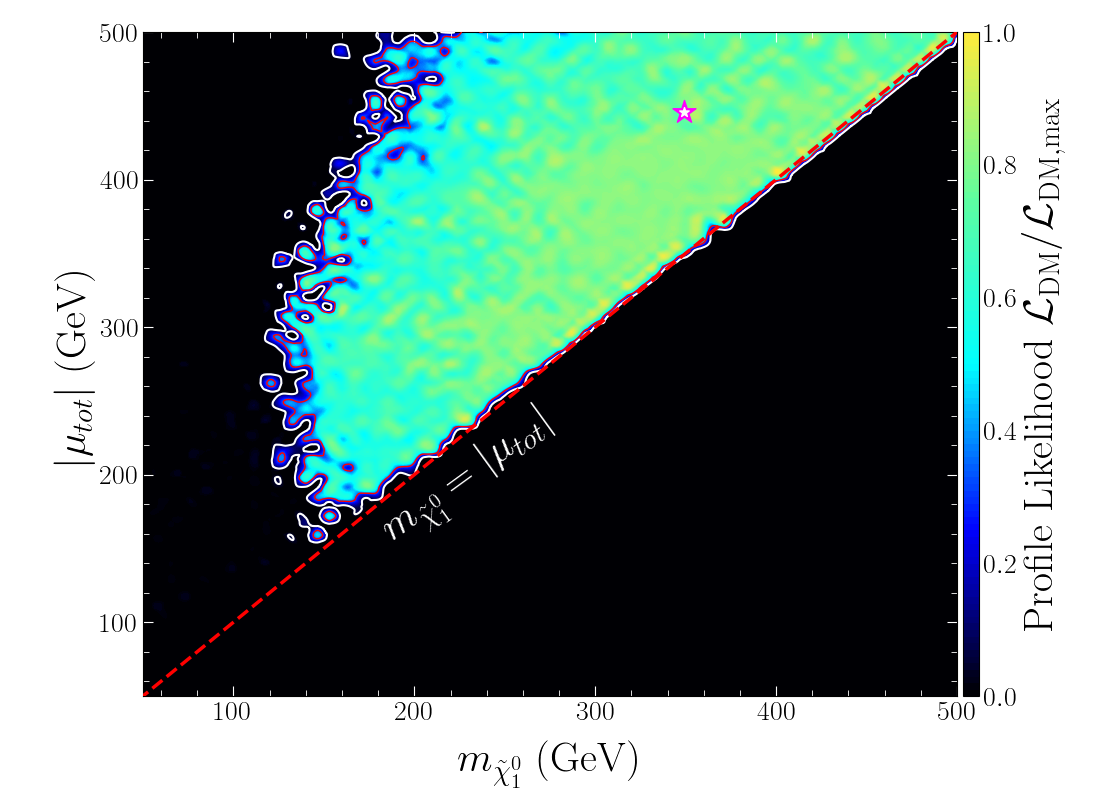}
        }

\caption{Two-dimensional profile likelihood of function $\mathcal{L}_{\rm DM}$ in Eq.~(\ref{PL-DM}) for $h \equiv h_1$ scenario, which is projected onto $|\kappa| - \lambda$ and $|\mu_{tot}|- m_{\tilde{\chi}_1^0}$ planes of $\mu$NMSSM. Since $\chi^2_{\rm Best} \simeq 0$ for the best point (marked with a star symbol in figure), the boundaries for $1 \sigma$ (red solid line) and $2\sigma$ (white solid line) confidence intervals correspond to $\chi^2 \simeq 2.3$ and $\chi^2 \simeq 6.18$, respectively. Dotted-dashed line represents relation $2{\kappa}=\lambda$. Points around $m_{\tilde{\chi}_1^0} \simeq |\mu_{tot}|$ on right-hand panel obtained correct abundance by co-annihilation.  This figure shows $\mu$NMSSM's parameter region that can well explain DM experiments.  \label{Fig1}}
\end{figure*}

\begin{figure*}[t]
		\centering
		\resizebox{0.98\textwidth}{!}{
        \includegraphics{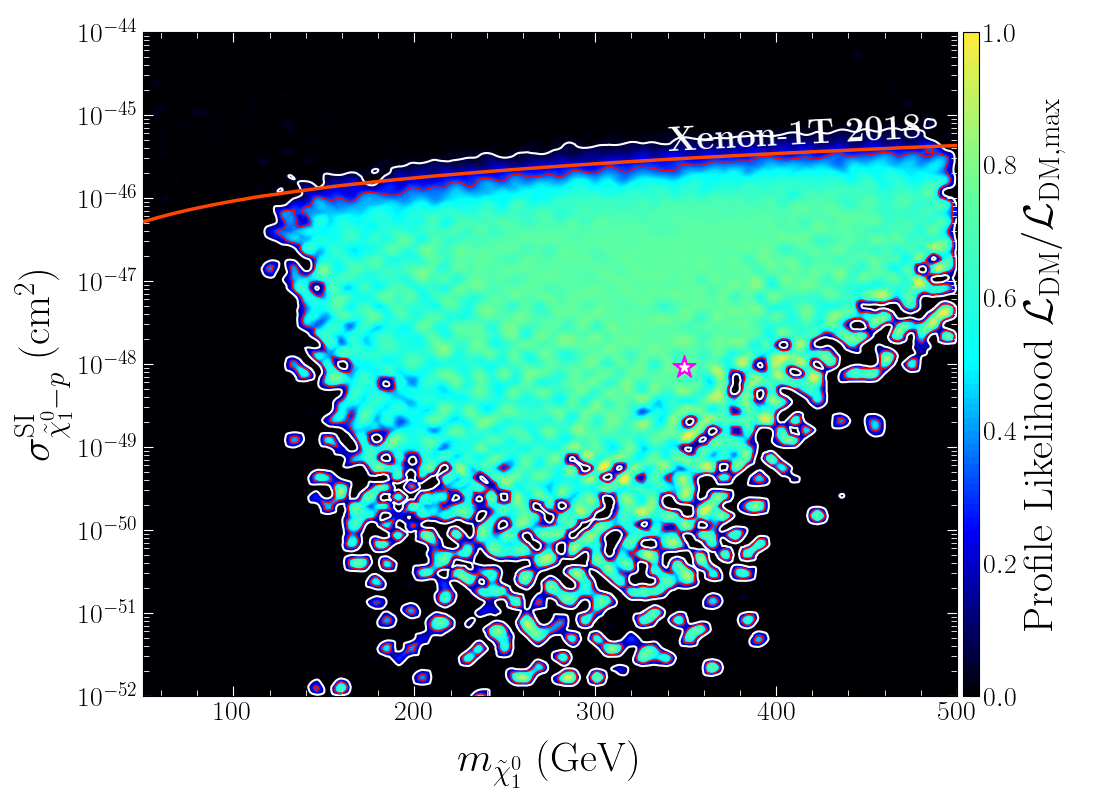}
        \includegraphics{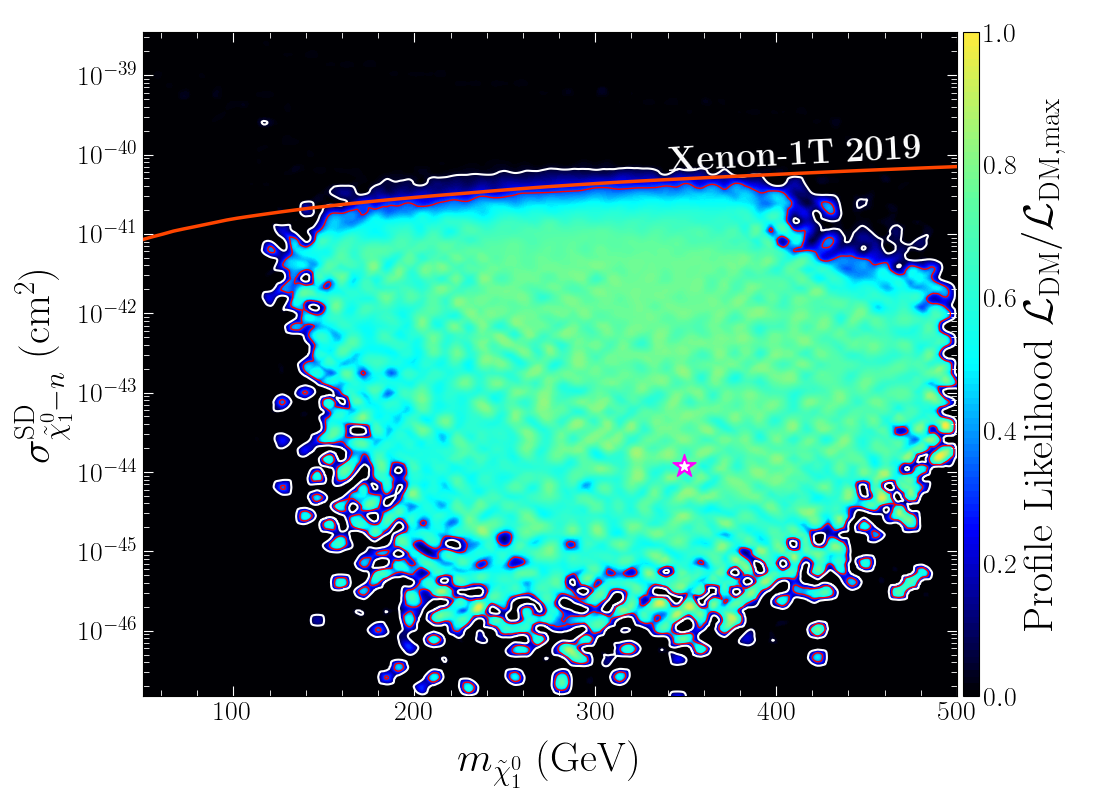}
        }

\caption{Same as Figure~\ref{Fig1}, but for profile likelihood projected onto $\sigma_{\tilde{\chi}_1^0-p}^{\rm SI} - m_{\tilde{\chi}_1^0}$ and  $\sigma_{\tilde{\chi}_1^0-n}^{\rm SD} - m_{\tilde{\chi}_1^0}$ plane. Red curve denotes latest XENON-1T's bound on SI and SD cross-section. \label{Fig2}}
\end{figure*}	

\subsection{Profile likelihood of DM physics}

The object of focus in this subsection is the theory's capability to explain the DM experiments. DM's two-dimensional (2D) PL is defined as
	\begin{eqnarray}
 \mathcal{L}_{\rm DM} (\Theta_A,\Theta_B)=\mathop{\max}_{\Theta_1,\cdots,\Theta_{A-1},\Theta_{A+1},\cdots, \Theta_{B-1}, \Theta_{B+1},\cdots} \left \{ \mathcal{L}_{\Omega h^2} \times \mathcal{L}_{DD} \times \mathcal{L}_{IDD} \right \} (\Theta),  \label{PL-DM}
	\end{eqnarray}
where $\Theta = (\Theta_1\equiv \lambda,\Theta_2 \equiv \kappa,\cdots)$ are the input parameters of $\mathcal{L}_{i}$ ($i = \Omega h^2, \cdots$), and the maximization is realized by varying the parameters other than $\Theta_A$ and $\Theta_B$. The $1\sigma$ ($2 \sigma$) confidence interval on the $\Theta_A - \Theta_B$ plane is determined by the criteria of $(\delta \chi^2 \equiv \chi^2 - \chi^2_{Best}) \leq 2.3$ ($\delta \chi^2 \leq 6.18$), where $\chi^2 \equiv -2 \ln {\mathcal{L}_{\rm DM}}(\Theta_A,\Theta_B)$ and $\chi^2_{Best}$ denotes the best point's $\chi^2$. Two subtleties about $\mathcal{L}_{\rm DM}(\Theta_A, \Theta_B)$ must be clarified. One is that it is completely different from the PL of the total likelihood function $\mathcal{L}$ in Eq.~(\ref{Likelihood}). Specifically, the $\chi^2$ of $\mathcal{L}$ is extremely dominated by the Higgs data fit, which contains 102 measurements in the code \textsf{HiggsSignal-2.2.3}~\cite{HiggsSignal}. Thus, studying the distribution of $\mathcal{L}(\Theta_A, \Theta_B)$ is roughly equivalent to performing a Higgs data fit on the $\Theta_A - \Theta_B$ plane, which has nothing to do with DM physics. The other subtlety is that some of the scanned samples may be in serious conflict with the Higgs data or B-physics measurements, although they coincide well with the DM experiments. Since $\mathcal{L}$ strongly disfavors these samples, they should be neglected. Therefore, only the following samples are considered in studying $\mathcal{L}_{\rm DM} (\Theta_A,\Theta_B)$.
\begin{itemize}
\item Its $p$ value in the Higgs data fit is larger than $0.05$, which implies that it coincides with the data at $95\%$ confidence level.
\item It is consistent with the extra Higgs search results implemented in the code \textsf{HiggsBounds-5.3.2}.
\item It accounts for the measurement of $Br(B \to X_s \gamma)$ and $Br(B_s^0 \to \mu^+ \mu^-)$ at $2 \sigma$ level~\cite{Tanabashi:2018oca}.
\end{itemize}

\begin{figure*}[t]
		\centering
		\resizebox{0.98\textwidth}{!}{
        \includegraphics{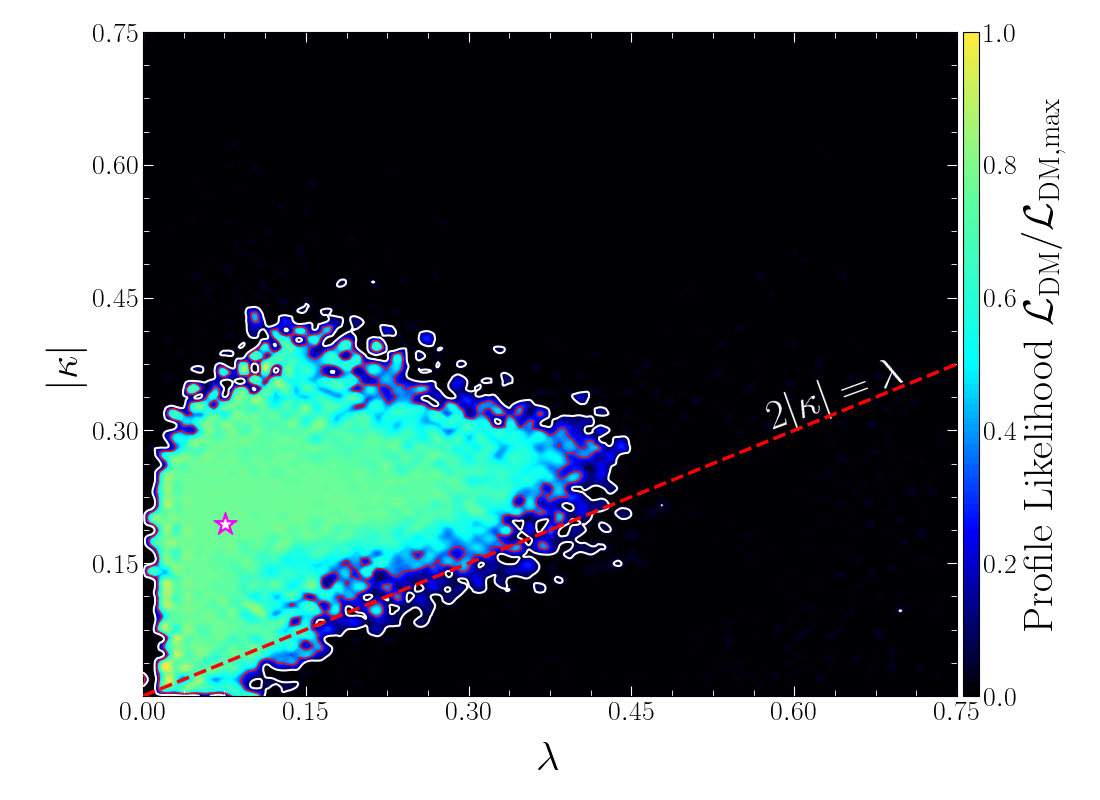}
        \includegraphics{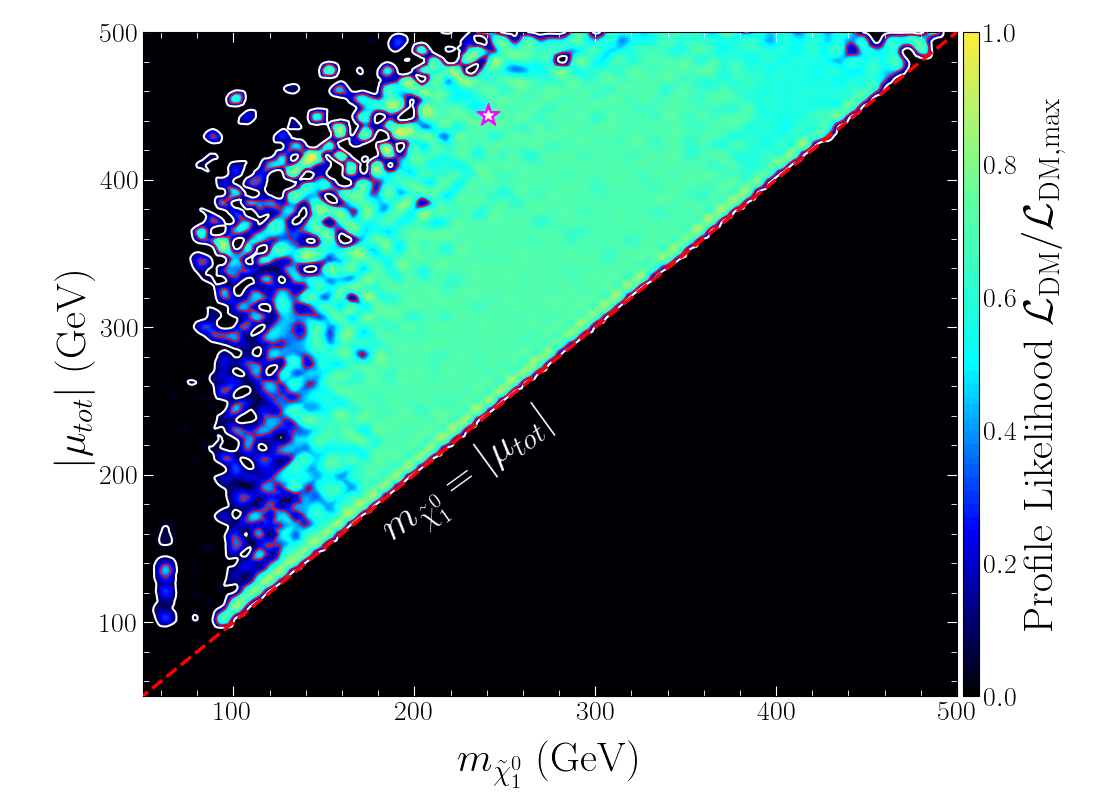}
        }

\caption{Same as Figure~\ref{Fig1}, but for $h \equiv h_2$ scenario.  \label{Fig3}}
\end{figure*}	

\begin{figure*}[t]
		\centering
		\resizebox{0.98\textwidth}{!}{
        \includegraphics{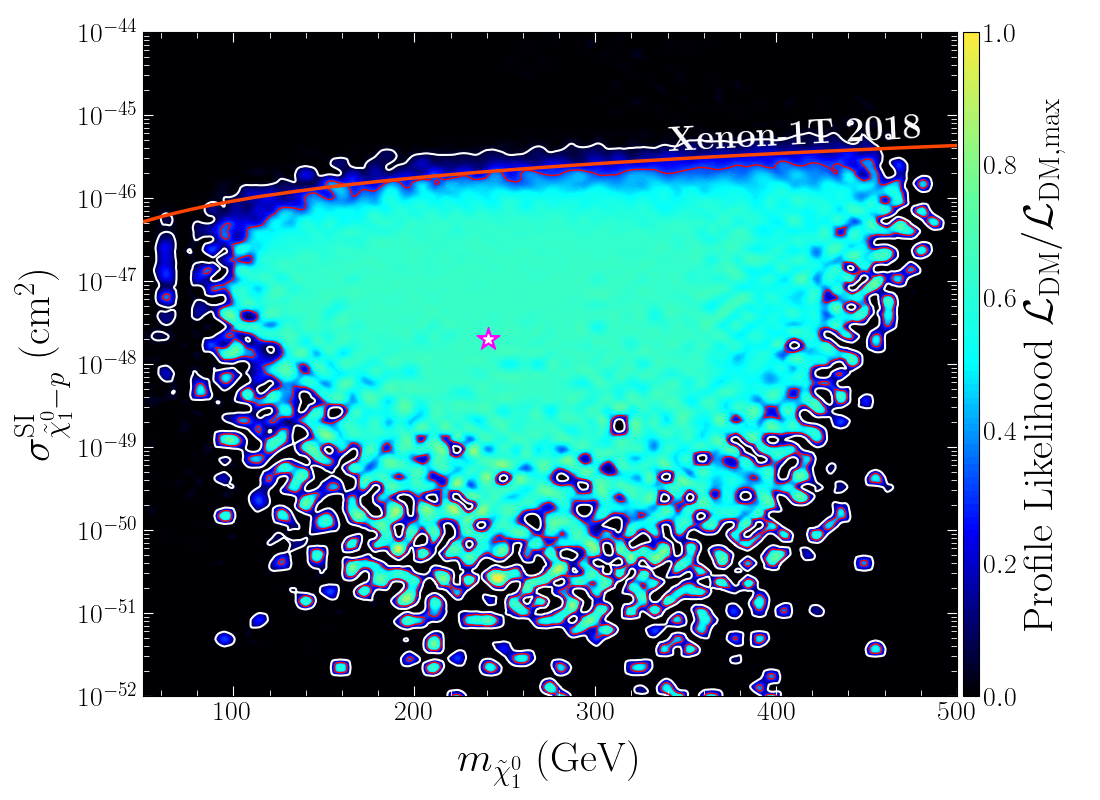}
        \includegraphics{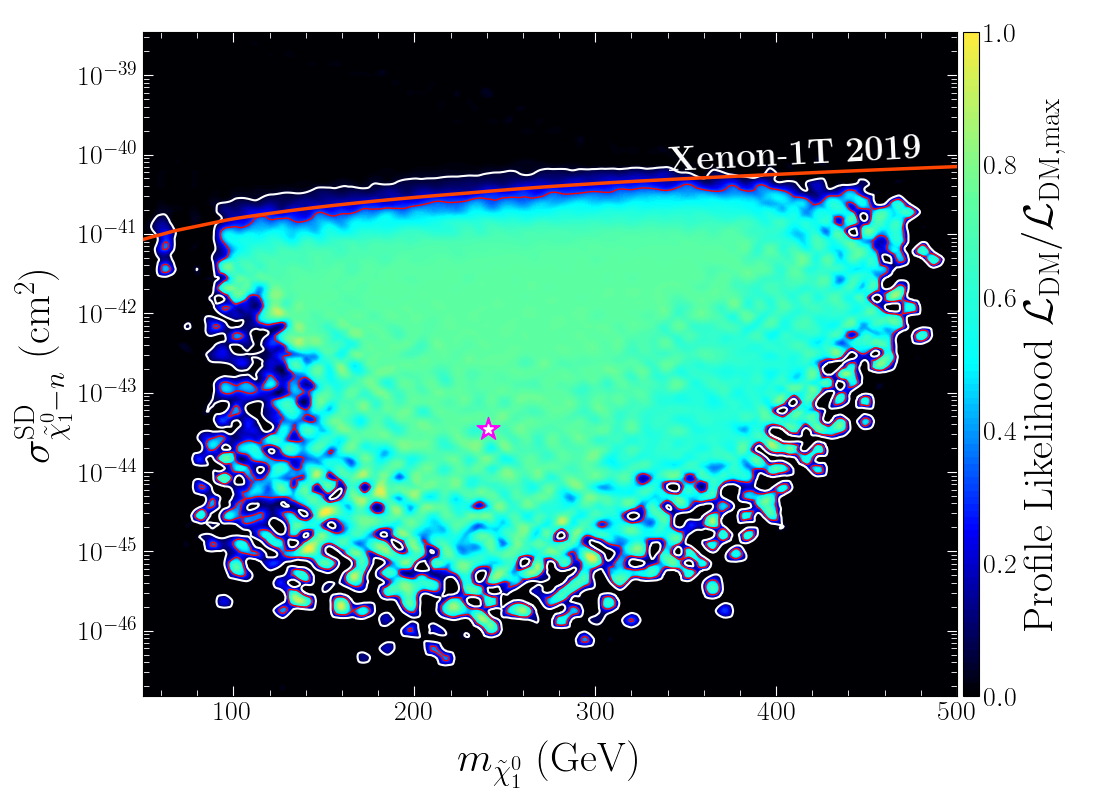}
        }

\caption{Same as Figure~\ref{Fig2}, but for $h \equiv h_2$ scenario.  \label{Fig4}}
\end{figure*}	

Using the refined samples, different 2D PL maps are shown in Figures \ref{Fig1} to \ref{Fig4}. Figure~\ref{Fig1} plots the credible interval (CI) on $|\kappa| - \lambda$ and $|\mu_{tot}|- m_{\tilde{\chi}_1^0}$ plane for the $h \equiv h_1$ scenario. This plot indicates that the theory can explain the DM experiments quite well when $ \lambda \lesssim 0.5$, $0.05 \lesssim |\kappa| \lesssim 0.5$, $m_{\tilde{\chi}_1^0} \gtrsim 140 ~{\rm GeV}$, and $|\mu_{tot}| \gtrsim 160~{\rm GeV}$. The best point locates at approximately $\lambda = 0.05$, $\kappa = 0.28$, $m_{\tilde{\chi}_1^0} = 350~{\rm GeV}$, and $\mu_{tot} = 450~{\rm GeV}$. Additionally, $2 |\kappa|/\lambda $ can be much larger than 1 to keep $\tilde{\chi}_1^0$ Singlino-dominated, which is one distinct feature of the $\mu$NMSSM. Figure~\ref{Fig2} shows the $1\sigma$ and $2\sigma$ CIs on the $\sigma_{\tilde{\chi}_1^0-p}^{\rm SI} - m_{\tilde{\chi}_1^0}$ and  $\sigma_{\tilde{\chi}_1^0-n}^{\rm SD} - m_{\tilde{\chi}_1^0}$ planes. This plot indicates that the SI and SD cross-sections can be as low as $10^{-51}~{\rm cm^2}$ and $10^{-46}~{\rm cm^2}$, respectively, over a broad mass range of $\tilde{\chi}_1^0$. This conclusion comes from the fact that $\lambda$ can be small and the ratio $m_{\tilde{\chi}_1^0}/\mu_{tot}$ in Eqs.(\ref{C_xsxshsm})-(\ref{C_xsxshsm}) can be much less than 1. Figures~\ref{Fig3} and ~\ref{Fig4} are similar to Figures~\ref{Fig1} and ~\ref{Fig2}, except for the $h \equiv h_2$ scenario. Compared with the $h \equiv h_1$ scenario, this scenario has the following different characteristics.
\begin{itemize}
\item The CI on $|\kappa| - \lambda$ plane shrinks significantly.
\item $m_{\tilde{\chi}_1^0}$ as low as $100~{\rm GeV}$ can explain the experiments.
\item The best point locates at approximately $\kappa = 0.2$ and $m_{\tilde{\chi}_1^0} = 250~{\rm GeV}$.
\end{itemize}
In addition to the fact that the parameter space of the $h \equiv h_2$ scenario is relatively narrow to fit well with the Higgs data~\cite{Cao:2012fz}, another reason for the difference is that, since $h_s = h_1 $ is light, a moderately light $\tilde{\chi}_1^0$ can make the annihilation $\tilde{\chi}_1^0 \tilde{\chi}_1^0 \to h_s A_s$ occur.

A detailed explanation of the PL figure, including how to plot it and understand it correctly, was presented in our previous works~\cite{Cao:2019qng,Cao:2019aam}. For the sake of brevity, it is not repeated here. Besides, the fact that the marginal posterior PDF and the PL differ significantly in their definitions is worth noting. Therefore, they are complementary to each other in describing the results of the present work.

\subsection{Constraints from LHC search for electroweakinos}

\begin{figure*}[t]
		\centering
		\resizebox{0.5\textwidth}{!}{
        \includegraphics{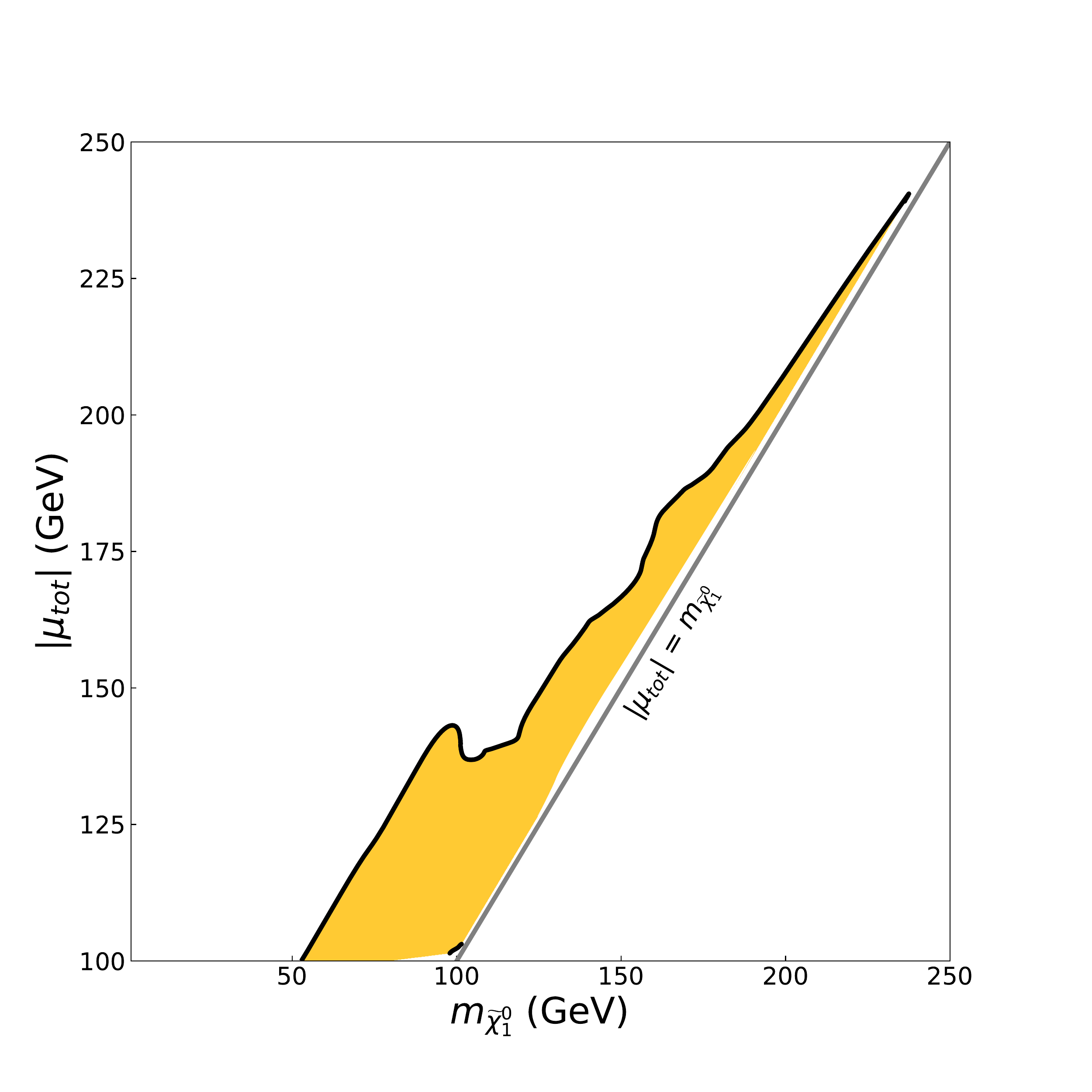}
        }
\caption{Constraint from latest ATLAS analysis of di-lepton signal at 13-TeV LHC, which is shown on $|\mu_{tot} | -m_{\tilde{\chi}_1^0}$ plane. Analysis is concentrated on compressed mass spectra~\cite{Aad:2019qnd} and excludes the yellow-shaded region at $95\%$ confidence level when $\lambda = 0.01$ and $\tan \beta = 10$ are fixed.     \label{Fig5}}
\end{figure*}

\begin{table}[]
\resizebox{1\textwidth}{!}
{
\begin{tabular}{|c|c|c|c|c|}
\hline
 &
  \multicolumn{3}{c|}{$\rm h \equiv h_1 $ Scenario} &
  $\rm h \equiv h_2 $ Scenario \\ \hline
 &
  Point $\rm \uppercase\expandafter{\romannumeral1}$ &
  Point $\rm \uppercase\expandafter{\romannumeral2}$ &
  Point $\rm \uppercase\expandafter{\romannumeral3}$ &
  Point $\rm \uppercase\expandafter{\romannumeral4}$ \\ \hline
$\lambda$ &
  0.026 &
  0.227 &
  0.038 &
  0.077 \\
$\kappa$ &
  0.236 &
  0.328 &
  -0.212 &
  0.199 \\
$\tan{\beta}$ &
  9.750 &
  6.698 &
  9.454 &
  8.092 \\
$\mu~(\rm GeV)$ &
  322.6 &
  243.8 &
  226.7 &
  291.3 \\
$\mu_{\rm tot}~(\rm GeV)$ &
  335.0 &
  331.9 &
  248.0 &
  338.2 \\
$A_{\kappa}~(\rm GeV)$ &
  -310.5 &
  -238.1 &
  21.58 &
  -415.8 \\
$A_t~(\rm GeV)$ &
  4244 &
  4112 &
  2999 &
  2959 \\ \hline
$m_{h_1}~(\rm GeV)$ &
  125.2 &
  125.2 &
  125.1 &
  93.67 \\
$m_{h_2}~(\rm GeV)$ &
  134.8 &
  203.9 &
  233.3 &
  125.2 \\
$m_{h_3}~(\rm GeV)$ &
  835.3 &
  1273 &
  794.6 &
  1053 \\
$m_{A_s}~(\rm GeV)$ &
  326.9 &
  310.6 &
  86.62 &
  388.4 \\
$m_{\tilde{\chi}_1^0}~(\rm GeV)$ &
  228.6 &
  249.2 &
  236.7 &
  255.4 \\
$m_{\tilde{\chi}_2^0}~(\rm GeV)$ &
  343.5 &
  343.2 &
  254.5 &
  386.7 \\
$m_{\tilde{\chi}_3^0}~(\rm GeV)$ &
  345.4 &
  345.5 &
  258.0 &
  348.7 \\
$m_{\tilde{\chi}_1^{\pm}}~(\rm GeV)$ &
  344.6 &
  340.7 &
  255.7 &
  387.8 \\ \hline
$V_{h}^{\rm NSM}, V_{h}^{\rm SM}, V_{h}^{\rm S}$ &
  -0.20,  -0.94,  0.30 &
  -0.30, -0.95, -0.06 &
  -0.21, -0.98, 0.02 &
  0.01, 0.99, 0.17 \\
$V_{h_s}^{\rm NSM}, V_{h_s}^{\rm SM}, V_{h_s}^{\rm S}$ &
  -0.05,  -0.29,  -0.96 &
  -0.05, 0.08, -0.99 &
  -0.02, -0.01, -0.99 &
  0.01, -0.17, 0.99 \\
$N_{13}, N_{14}, N_{15}$ &
  0.01, -0.02, 0.99 &
  0.14, -0.21, 0.97 &
  -0.17, -0.18, -0.97 &
  0.04, -0.07, 0.99 \\
$N_{23}, N_{24}, N_{25}$ &
  -0.71, 0.71, 0.02 &
  0.71, 0.71, 0.05 &
  -0.71, 0.71, 0.05 &
  -0.71, 0.70, 0.08 \\
$N_{33}, N_{34}, N_{35}$ &
  0.71, 0.71, 0.01 &
  -0.69, 0.68, 0.25 &
  -0.69, -0.69, -0.25 &
  0.71, 0.71, 0.02 \\ \hline
$\Omega h^2$ &
  0.125 &
  0.125 &
  0.131 &
  0.124 \\
$\sigma^{SI}_{\tilde{\chi}_1^0 - p}(\rm cm^2)$ &
  $2.83 \times 10^{-48}$ &
  $2.59 \times 10^{-47}$ &
  $1.01 \times 10^{-47}$ &
  $6.36 \times 10^{-48}$ \\
$\sigma^{SD}_{\tilde{\chi}_1^0 - n}(\rm cm^2)$ &
  $2.47 \times 10^{-46}$ &
  $1.95 \times 10^{-41}$ &
  $5.49 \times 10^{-43}$ &
  $2.50 \times 10^{-44}$ \\
Annihilation processes &
  \begin{tabular}[c]{@{}c@{}}$\quad$\\ $\;\;\;    71\% \rm \;\; \tilde{\chi}_1^0 \tilde{\chi}_1^0 \; \to \; h_s \; A_s$ \\ $\;\;\; 20\%  \rm \;\;    \tilde{\chi}_1^0 \tilde{\chi}_1^0 \; \to \; h \;\;\, A_s $\\ $\;\; 7.4\% \rm \;\; \tilde{\chi}_1^0 \tilde{\chi}_1^0 \; \to \; h_s \; h_s$\\ $ 1.3\% \rm  \; \; \tilde{\chi}_1^0 \tilde{\chi}_1^0 \; \to \; h \; h_s$\\ ......\end{tabular} &
  \begin{tabular}[c]{@{}c@{}}$\quad$\\ $34\% \rm \quad  \tilde{\chi}_1^0 \tilde{\chi}_1^0 \; \to \; t \; \bar{t}$ \\ $\quad 33\% \rm \quad  \tilde{\chi}_1^0 \tilde{\chi}_1^0 \; \to \; h_s \; A_s $\\ $\quad 27\% \rm \quad  \tilde{\chi}_1^0 \tilde{\chi}_1^0 \; \to \; h_s \; h_s$\\ $\;\; 1.4\% \rm  \quad  \tilde{\chi}_1^0 \tilde{\chi}_1^0 \; \to \; h \; A_s$\\ $\qquad \  1.1\%  \rm \quad  \tilde{\chi}_1^0 \tilde{\chi}_1^0 \; \to \; W^+ \; W^-$\\ ......\end{tabular} &
  \begin{tabular}[c]{@{}c@{}}$\quad$\\ $5.4\% \rm \quad  \tilde{\chi}_2^0 \tilde{\chi}_1^\pm \; \to \; d \; \bar{u}$ \\ $5.4\% \rm \quad    \tilde{\chi}_2^0 \tilde{\chi}_1^\pm \; \to \; s \; \bar{c} $\\ $4.0\%   \rm \quad  \tilde{\chi}_2^0 \tilde{\chi}_1^\pm \; \to \; b \; \bar{t}$\\ $4.0\%  \rm \quad  \tilde{\chi}_1^0 \tilde{\chi}_1^\pm \; \to \; s \; \bar{c}$\\  $4.0\% \rm  \quad   \tilde{\chi}_1^0  \tilde{\chi}_1^\pm \, \to \; d \; \bar{u} $\\ ......\end{tabular} &
  \multicolumn{1}{l|}{\begin{tabular}[c]{@{}l@{}}$96\% \rm \quad  \tilde{\chi}_1^0 \tilde{\chi}_1^0 \; \to \; h_s \; A_s$ \\ $3.4\% \rm  \;\;\;  \tilde{\chi}_1^0 \tilde{\chi}_1^0 \; \to \; h_s \; h_s$\end{tabular}} \\
$\tilde{\chi}_2^0$ Decay channels &
  $100\%\rm  \quad  \tilde{\chi}_2^0  \; \to \; \tilde{\chi}_1^0 \; Z$ &
  $100\% \rm \quad  \tilde{\chi}_2^0  \; \to \; \tilde{\chi}_1^0 \; Z$ &
  \begin{tabular}[l]{@{}l@{}}$\quad$\\ $16\%\rm \quad  \tilde{\chi}_2^0  \; \to \; \tilde{\chi}_1^0 \; d \; \bar{d} \;(\, d=d,\; s \,)$\\ $12\%\rm  \quad  \tilde{\chi}_2^0  \; \to \; \tilde{\chi}_1^0 \; u \; \bar{u} (\; u=u,\; c \;)$\\ $10\% \rm \quad  \tilde{\chi}_2^0  \; \to \; \tilde{\chi}_1^0 \; b \; \bar{b} $\\ $7.4\%  \rm  \quad  \tilde{\chi}_2^0  \; \to \; \tilde{\chi}_1^0 \; \nu \; \bar{\nu}\; (\, \nu= \nu_e,\;\nu_{\mu}, \; \nu_{\tau}\,)$\\ $3.7\% \rm \quad  \tilde{\chi}_2^0  \; \to \; \tilde{\chi}_1^0 \; l \; \bar{l} (\; l=e,\; \mu \;)$\\ $3.5\% \rm \quad  \tilde{\chi}_2^0  \; \to \; \tilde{\chi}_1^0 \; \tau \; \bar{\tau} \;$\\ $\quad$\end{tabular} &
   \begin{tabular}[c]{@{}c@{}}$57\% \rm \quad  \tilde{\chi}_2^0  \; \to \; \tilde{\chi}_1^0 \; Z$\\ $43\% \rm \quad  \tilde{\chi}_2^0  \; \to \; \tilde{\chi}_1^0 \; h_s$\end{tabular} \\
\multicolumn{1}{|c|}{$\tilde{\chi}_3^0$ Decay channels} &
  $100\%\rm  \quad  \tilde{\chi}_3^0  \; \to \; \tilde{\chi}_1^0 \; Z$ &
  $100\%\rm  \quad  \tilde{\chi}_3^0  \; \to \; \tilde{\chi}_1^0 \; Z$ &
  \begin{tabular}[c]{@{}l@{}}$\quad$\\ $11\% \rm \quad\;  \tilde{\chi}_3^0  \; \to \; \tilde{\chi}_1^0 \; d \; \bar{d}\;(\, d=d,\; s \,)$\\ $8.3\% \rm \quad  \tilde{\chi}_3^0  \; \to \; \tilde{\chi}_1^0 \; u \; \bar{u}$\\ $7.9\% \rm \quad  \tilde{\chi}_3^0  \; \to \; \tilde{\chi}_1^0 \; c \; \bar{c}$\\ $6.8\% \rm \quad  \tilde{\chi}_3^0  \; \to \; \tilde{\chi}_1^0 \; b \; \bar{b}$\\ $5.6\% \rm \quad  \tilde{\chi}_3^0  \; \to \; \tilde{\chi}_3^0 \; b \; \bar{b} $\\ $5.0\%   \quad  \tilde{\chi}_3^0  \; \to \; \tilde{\chi}_1^0 \; \nu \; \bar{\nu}\; (\, \nu= \nu_e,\;\nu_{\mu}, \; \nu_{\tau}\,)$\\ \qquad\qquad\qquad......\end{tabular} &
  \begin{tabular}[c]{@{}c@{}}$99.7\%\rm  \quad  \tilde{\chi}_3^0  \; \to \; \tilde{\chi}_1^0 \; Z$\\ $0.3\%\rm  \quad  \tilde{\chi}_3^0  \; \to \; \tilde{\chi}_1^0 \; h_s$\end{tabular} \\
\multicolumn{1}{|c|}{$\tilde{\chi}_1^{\pm}$  Decay channels} &
  $~100\% \rm \quad  \tilde{\chi}_1^\pm  \; \to \; \tilde{\chi}_1^0 \; W^\pm$ &
  $~100\% \rm \quad  \tilde{\chi}_1^\pm  \; \to \; \tilde{\chi}_1^0 \; W^\pm$ &
  \begin{tabular}[c]{@{}l@{}}$\quad$\\ $~33\% \rm \quad  \tilde{\chi}_2^\pm  \; \to \; \tilde{\chi}_1^0 \; u \; \bar{d} (\, u \; \bar{d}=u\; \bar{d}, \;  c\;\bar{ s} \,)$\\ $~11\% \rm \quad  \tilde{\chi}_1^\pm  \; \to \; \tilde{\chi}_1^0 \; \nu \; \bar{l}\; (\,\nu \; \bar{l}=\nu_e \; \bar{e} , \; \nu_{\nu}\; \bar{\mu}, \;\nu_{\tau} \; \bar{\tau} \,)$\\ $\quad$\end{tabular} &
  $~100\% \rm \quad  \tilde{\chi}_1^\pm  \; \to \; \tilde{\chi}_1^0 \; W^\pm$ \\ \hline
\end{tabular}
}
\caption{Detailed information of four benchmark points that agree well with all considered experiments. Number before each annihilation process represents its fraction in contributing to total DM annihilation cross-section at freeze-out temperature. Number before each decay denotes its branching ratio. }\label{tab:tab4}
\end{table}

In the GNMSSM, the natural electroweak symmetry breaking tends to a Higgsino mass of several hundreds of GeV. In this case, the LHC will produce Higgsino pair events copiously, and the Higgsino’s property is strongly restricted by searching for multi-lepton signals. Up to now, experimental analyses in this aspect usually considered Wino pair production and provided the Wino mass bound as a function of $m_{\tilde{\chi}_1^0}$ in the simplified model. Recently, the ATLAS collaboration analyzed $139 {\rm fb^{-1}}$ proton-proton collision data collected at the LHC with $\sqrt{s} = 13 {\rm TeV}$ and concluded that the LHC had already explored the region with the Wino mass up to approximately $700 {\rm GeV}$ and $m_{\tilde{\chi}_1^0}$ up to $300 {\rm GeV}$~\cite{Aad:2019vnb,Aad:2019vvi,ATLAS:2020ckz}. These analyses were applied to the Higgsino pair production process in the present study by elaborate Monte Carlo simulations and it was found that they can effectively limit the Higgsino’s property when $m_{\tilde{\chi}_1^0} \lesssim 100~{\rm GeV}$\footnote{A similar conclusion was obtained as shown in Figure 3 of~\cite{Cao:2019qng}, where the results on the upper panel of Figure 8 in~\cite{Sirunyan:2018ubx} are re-interpreted  in terms of the Higgsino pair production process. The latter result was obtained by a combined analysis of the multi-lepton signal by the CMS collaboration. Its exclusion capability on the Wino mass versus DM mass plane is roughly identical to that of the analyses in~\cite{Aad:2019vnb,Aad:2019vvi,ATLAS:2020ckz}. }. As a specific application, the $h$-funnel region in the $h \equiv h_2$ scenario has been excluded by the analyses. This result is consistent with our previous observation in~\cite{Cao:2018rix} that the area has been ruled out by CMS’s search for the multi-lepton signal at the 13-TeV LHC with  $35.9 {\rm fb^{-1}}$ data~\cite{Sirunyan:2018ubx}.

Above discussion reveals that the analyses in~\cite{Aad:2019vnb,Aad:2019vvi,ATLAS:2020ckz,Sirunyan:2018ubx} have little restriction on the theory since $\tilde{\chi}_1^0$ in the GNMSSM is preferred heavier than 100 GeV. This fact motivates us to consider an experimental analysis that can detect high-DM-mass region~\cite{Aad:2019qnd}. The analysis aims to explore compressed mass spectra and utilizes the $139 {\rm fb^{-1}}$ data of the LHC. Events with missing transverse momentum and two same-flavor, oppositely charged, low-transverse-momentum leptons are selected, and are further categorized by the presence of hadronic activity from initial-state radiation. The analysis was reproduced in the present work by using the simulation tools MadGraph5\_aMC@NLO-2.6.6~\cite{mad-1,mad-2} to generate the parton-level events of the processes $ p p \to \tilde{\chi}_2^0 \tilde{\chi}_3^0 (j), \tilde{\chi}_2^0 \tilde{\chi}_1^\pm (j), \tilde{\chi}_3^0 \tilde{\chi}_1^\pm (j)$, and $\tilde{\chi}_1^\pm \tilde{\chi}_1^\mp (j)$; Pythia-8.2~\cite{pythia} for parton fragmentation and hadronization; Delphes-3.4.2~\cite{delphes} for fast simulation of the performance of the ATLAS detector; and CheckMATE-2.0.26~\cite{cmate-1,cmate-2,cmate-3} to implement the analysis' cut selections. The validation of the method is provided in our recent work~\cite{cao:2021new-work}.

In Figure~\ref{Fig5}, the excluded region is shown on the $|\mu_{tot} | -m_{\tilde{\chi}_1^0}$ plane. $\lambda = 0.01$, $\tan \beta = 10$, and $M_1 = M_2 = 3~{\rm TeV}$ are fixed, and the procedure depicted in~\cite{cao2019bayesian} is followed to plot the figure. The result indicates that the analysis can explore $m_{\tilde{\chi}_1^0}$ up to $230~{\rm GeV}$ and significantly impact the co-annihilation mechanism. In addition, it is noticeable that the excluded region in the present work is broader than that on the upper panel of Figure~14 in~\cite{Aad:2019qnd}.  This is because $\tilde{\chi}_1^0$ is Singlino-dominated in this work, instead of Higgsino-dominated in~\cite{Aad:2019qnd}, so more processes contribute to the signal.

To fully demonstrate the Singlino-dominated DM scenario's characteristics, the details of four benchmark points are provided in Table~\ref{tab:tab4}. The first three points belong to the $h \equiv h_1$ scenario and the last one is from the $h \equiv h_2$ scenario. They all predict $m_{\tilde{\chi}_1^0} \gtrsim 230 {\rm GeV}$ and thus are difficult to detect at the LHC. In addition to the characteristics listed in the table, the following features merit emphasizing.
\begin{itemize}
\item If the mass splitting between $\tilde{\chi}_{i}^0$ ($i=2,3$) and $\tilde{\chi}_1^0$ is larger than any of the neutral Higgs boson masses, $\tilde{\chi}_{i}^0$ will decay with sizable branching ratios into the Higgs bosons. Otherwise, it will decay into a real or virtual $Z$ boson.
\item Although the process $\tilde{\chi}_1^0 \tilde{\chi}_1^0 \to h_s A_s$ is mainly responsible for the abundance in most cases, its role diminishes when $m_{h_s} + m_{A_s} $ approaches $2 m_{\tilde{\chi}_1^0}$ from below. This situation is demonstrated by Points 1 and 2.
\item Since $V_h^S = 0.3$ is sizable, $\tilde{\chi}_1^0 \tilde{\chi}_1^0 \to h A_s$ for Point 1 also plays a crucial role in determining the abundance.
\item Since $\lambda = 0.227$ is sizable, $\sigma^{\rm SI}_{\tilde{\chi}_1^0-p}$ and $\sigma^{\rm SD}_{\tilde{\chi}_1^0-n}$ for Point 2 are relatively large.
\item Since $m_{\tilde{\chi}_1^0}/\mu_{tot} = 0.95$ approaches 1,  $\sigma^{\rm SI}_{\tilde{\chi}_1^0-p}$ and $\sigma^{\rm SD}_{\tilde{\chi}_1^0-n}$ for Point 3 are relatively large even though $\lambda$ is small.
\item Point 4 predicts a significant cancellation between the $h$- and $h_s$-mediated contributions to $\sigma^{\rm SI}_{\tilde{\chi}_1^0-p}$.
\end{itemize}

\section{Conclusions}

Motivated by the increasingly tight limitation of the direct detection experiments on traditional neutralino DM in the natural MSSM and $Z_3$-NMSSM, the $Z_3$-NMSSM is extended in the present work by adding a $\mu \hat{H}_u \cdot \hat{H}_d $ term in its superpotential, and whether the Singlino-dominated neutralino can act as a feasible DM candidate is studied. Different from the $Z_3$-NMSSM, the extended theory describes the neutralino's property by four independent parameters: $\lambda$, $\mu_{tot}$, $m_{\tilde{\chi}_1^0}$, and $\kappa$. The first three parameters strongly influence the DM-nucleon scattering rate, while $\kappa$ usually only slightly affects the scattering. This characteristic implies that singlet-dominated particles may form a secluded DM sector. Under such a theoretical structure, the Singlino-dominated neutralino achieves the correct abundance by annihilating into a pair of singlet-dominated Higgs bosons by adjusting $\kappa$'s value. Its scattering with nucleons is suppressed when $\lambda v/\mu_{tot}$ is small.

Our speculations are verified by numerical results. Specifically, a likelihood function containing the current experimental and theoretical situations of DM physics, Higgs physics, and B physics was constructed, which was then utilized to guide sophisticated scans of the theory's parameter space by the nest sampling algorithm. The scanned samples were analyzed by several statistical quantities, such as the marginal posterior PDF and profile likelihood, to reveal the underlying physics. The results show that, in the theory's natural space for electroweak symmetry breaking, the DM obtains the correct abundance by $\tilde{\chi}_1^0 \tilde{\chi}_1^0 \to h_s A_s$ in most cases, and the SI and SD cross-sections may be as low as $10^{-51} {\rm cm^2}$ and $10^{-46} {\rm cm^2}$, respectively. Furthermore, the LHC search for electroweakinos restricts the theory very weakly.

It is emphasized herein that the $Z_3$-NMSSM differs significantly from the extended theory in at least three aspects. First, $m_{\tilde{\chi}_1^0}$ and $\kappa$ are no longer independent once one takes $\lambda$ and $\mu_{tot}$ as inputs to study the DM's property. Second, $|\kappa|$ must be less than $\lambda/2$ to keep the lightest neutralino Singlino-dominated. Its magnitude should be small after considering the tight constraints from the direct detection experiments. Third, due to the absence of the $\mu$-induced contributions to the Higgs squared mass, the singlet-dominated Higgs particles tend to be more massive. Combined with these facts, the DM cannot obtain the
correct abundance by $\tilde{\chi}_1^0 \tilde{\chi}_1^0 \to h_s A_s$. Instead, it must co-annihilate with the Higgsinos to keep consistent with the DM experimental results, which corresponds to the correlated parameter space $ \lambda \simeq 2 |\kappa| $ with $\lambda \lesssim 0.1$. The Bayesian evidence of the two theories was compared and it was found that their Jeffrey's-scale value is 8.05. This result implies that the considered experiments strongly prefer the extended theory. Since it resurrects the $Z_3$-NMSSM's broad parameter space that has been experimentally excluded, the extended theory is attractive and worthy of a careful study.

\section*{Acknowledgements}

We thank Di Zhang for useful discussions about the LHC search for supersymmetry. This work is supported by the National Natural Science Foundation of China (NNSFC) under Grant Nos. 11575053 and 12075076.



\providecommand{\href}[2]{#2}\begingroup\raggedright\endgroup

\end{document}